\documentclass[twocolumn]{aastex631}

\shortauthors{M. De Furio et al.}
\shorttitle{A-type Binaries with long-baseline interferometry}
\graphicspath{{./}{figures/}}
\begin{document}
\title{The Small Separation A-Star Companion Population: First Results with CHARA/MIRC-X}

\correspondingauthor{Matthew De Furio}
\email{defurio@umich.edu}

\author[0000-0003-1863-4960]{Matthew De Furio}
\affiliation{Department of Astronomy, University of Michigan, Ann Arbor, MI}

\author{Tyler Gardner}
\affiliation{Department of Astronomy, University of Michigan, Ann Arbor, MI}

\author{John Monnier}
\affiliation{Department of Astronomy, University of Michigan, Ann Arbor, MI}

\author{Michael R. Meyer}
\affiliation{Department of Astronomy, University of Michigan, Ann Arbor, MI}

\author{Kaitlin Kratter}
\affiliation{Department of Astronomy and Steward Observatory, University of Arizona, Tucson, AZ}

\author{Gail Schaefer}
\affiliation{CHARA Array, Georgia State University, Atlanta, GA 30302, USA}

\author{Narsireddy Anugu}
\affiliation{Department of Astronomy and Steward Observatory, University of Arizona, Tucson, AZ}
\affiliation{CHARA Array, Georgia State University, Atlanta, GA 30302, USA}

\author{Claire L. Davies}
\affiliation{Astrophysics Group, Department of Physics and Astronomy, University of Exeter, Stocker Road, Exeter, EX4 4QL, UK}

\author{Stefan Kraus}
\affiliation{Astrophysics Group, Department of Physics and Astronomy, University of Exeter, Stocker Road, Exeter, EX4 4QL, UK}

\author{Cyprien Lanthermann}
\affiliation{CHARA Array, Georgia State University, Atlanta, GA 30302, USA}

\author{Jean-Baptiste Le Bouquin}
\affiliation{Université Grenoble Alpes, CNRS, IPAG, 38000, Grenoble, France}

\author{Jacob Ennis}
\affiliation{Department of Astronomy, University of Michigan, Ann Arbor, MI}



\begin{abstract}

We present preliminary results from our long-baseline interferometry (LBI) survey to constrain the multiplicity properties of intermediate-mass A-type stars within 80pc.  Previous multiplicity studies of nearby stars exhibit orbital separation distributions well-fitted with a log-normal with peaks $>$15au, increasing with primary mass. The A-star multiplicity survey of \citet{DeRosa2014}, sensitive beyond 30au but incomplete below 100 au, found a log-normal peak around 390au. Radial velocity surveys of slowly-rotating, chemically peculiar Am stars identified a significant number of very close companions with periods $\leq$ 5 days, $\sim$ 0.1au, a result similar to surveys of O- and B-type primaries.  With the improved performance of LBI techniques, we can probe these close separations for normal A-type stars where other surveys are incomplete.  Our initial sample consists of 27 A-type primaries with estimated masses between 1.44-2.49M$_{\odot}$ and ages 10-790Myr, which we observed with the MIRC-X instrument at the CHARA Array.  We use the open source software CANDID to detect five companions, three of which are new, and derive a companion frequency of 0.19$^{+0.11}_{-0.06}$ over mass ratios 0.25-1.0 and projected separations 0.288-5.481 au.  We find a probability of 10$^{-6}$ that our results are consistent with extrapolations based on previous models of the A-star companion population, over mass ratios and separations sampled. Our results show the need to explore these very close separations to inform our understanding of stellar formation and evolution processes.
\end{abstract}

%\keywords{editorials, notices --- 
%miscellaneous --- catalogs --- surveys}


\section{Introduction} \label{sec:intro}
Multiplicity is a common outcome of the star formation process, the properties of which, e.g. orbital separation and mass ratio, seemingly depend on the primary mass of the system \citep{DucheneKraus2013,Moe2017}. Multiple systems tend to form through two common processes: disk fragmentation \citep{Adams1989, Bonnell1994, Kratter2008} and turbulent fragmentation \citep{Goodwin2004, Offner2010}.  Disk fragmentation is thought to account for the origin of companions at separations on the order of the size of the disk (10-100s au), while turbulent fragmentation is thought to account for the origin of companions at wider separations (100s-1000s au).  Close companions (0.1 - 10 au) are found around various types of stars regardless of mass \citep[e.g.][]{Reid2006, Raghavan2010, Sana2012}, but additional processes are needed to explain their separations.  Orbits can shrink, even for initially wide binaries, through interactions between the companion and infalling gas from the natal cloud, direct interactions with a circumstellar disk, and dynamical interactions of unstable multiple systems \citep{Bate2002MNRAS.336..705B, Bate2003MNRAS.339..577B, Offner2010, Bate2012MNRAS.419.3115B}.  These small separation companions have significant impacts on our fundamental understanding of planet formation and stellar evolution, as well as star formation.

%Small separation ($\lesssim$ 0.1 au) but their formation cannot be explained through these standard mechanisms and require additional processes of orbital evolution to account for their existence \citep{MoeKratter2018}.

Many volume-limited surveys have characterized the stellar multiple population based on primary star mass, mass ratio, and orbital separation. They find that the companion orbital separation distribution can be fitted as a log-normal distribution with a peak $>$ 15 au that increases with primary mass. Low-mass M-type primaries have a companion separation peak around 20 au \citep{Janson2012, Winters2019}, solar-type primaries have a peak at 50 au \citep{Raghavan2010}, and A-type primaries have a peak around 390 au \citep{DeRosa2014}. However, the A-type multiplicity survey of \citet{DeRosa2014} is incomplete for separations $<$ 100 au, leaving open the possibility for a significant population of close separation companions to A-type primaries.  

Radial velocity surveys of chemically peculiar Am stars with small rotational velocities identified a significant number of very close companions with a characteristic period of 5 days, $\sim$ 0.1 au \citep{AbtLevy1985,Carquillat2007}.  Close companions could drive the decreased rotational velocity and thus the peculiar chemical composition of Am stars.
But, a similar peak has been identified for O-type primaries \citep{Sana2012}.  Interferometric studies of OB primaries in the Orion Nebula Cluster found a bi-modal distribution with peaks at $\sim$ 1 au and 400-600 au \citep{Karl2018}.   A similar population of close companions to normal A-stars would be vitally important to understanding possible planetary architectures around intermediate-mass stars and the formation mechanisms required to form small separation binaries as a function of primary mass.  

If small separation companions to intermediate mass stars are common, they could represent a likely formation mechanism of Type Ia supernova progenitors, i.e. white dwarfs with masses $\geq$ 0.7 M$_{\odot}$ evolved from intermediate mass stars \citep{Cummings2018ApJ...866...21C}.  One such binary with a high mass ratio could result in the double-degenerate scenario where two white dwarfs with masses $\sim$ 0.7-0.8 M$_{\odot}$ form and eventually merge, producing a Type Ia Supernova as its combined mass exceeds the Chandrashekhar mass limit \citep{Webbink1984ApJ...277..355W}.  These types of multiples could also produce the single-degenerate scenario where a white dwarf accretes hydrogen from a close companion star and eventually detonates when surpassing the Chandrashekhar mass limit \citep{Whelan1973ApJ...186.1007W}.

Identifying close companions to A-stars with the radial velocity method can be more difficult than to solar-type stars as their spectral lines are quite broad due to their fast rotation \citep{Borgniet2019}.  Radial velocity surveys also require frequent observations over multiple years to detect companions at several au in separation for a typical A-type star.  Extreme adaptive optics (AO) systems can achieve angular resolutions down to $\sim$ 20 milli-arcseconds (mas), but cannot resolve the region in which Am stars have close companions ($\sim$ 0.1 au) at distances $>$ 5 pc. Therefore, we used the Michigan Infra-Red Combiner-eXeter (MIRC-X) instrument \citep{Anugu2020} on the long-baseline interferometric array at the Center for High Angular Resolution Astronomy \citep[CHARA, ][]{tenBrummelaar2005ApJ...628..453T} to observe a pilot sample of A-type stars within 80 pc and search for close companions down to separations of 0.5 mas.

In Section 2, we describe the data and the methods to identify companions. In Section 3, we present the companion detections, describe our detection limits, and make a preliminary estimate of the close companion population of A-type primary stars. In Section 4, we compare our results to various multiplicity surveys and discuss the implications. In Section 5, we summarize our conclusions.

\section{Methods} \label{sec:methods}

\subsection{Observations} \label{subsec:observations}
We observed 26 A-type stars over the course of two nights (UT December 20-21, 2020) at the CHARA Array with the MIRC-X instrument in the near-infrared H photometric band ($\sim$ 1.6 microns), listed in Table \ref{table1}.  The CHARA Array is an interferometer made up of six 1-meter telescopes which operate in the optical and near infrared with baselines ranging from 34-331 meters.  MIRC-X is a beam combiner instrument \citep{Monnier2006SPIE.6268E..1PM, Anugu2020} that operates in the near infrared, using all CHARA telescopes.  All of our observations were taken in the grism mode with R $\sim$ 190.  This spectral mode allows us to search out to wide angular scales ($\sim$ 0.3") as the increased spectral dispersion increases the interferometric field of view \citep{Anugu2020}.  At the same time, the spectral resolution is not so high that flux is spread out across many spectral channels, reducing the signal-to-noise (S/N) and thus reducing the achievable contrast.  The R190 grism is ideal for observing our relatively bright targets (typically H = 4-5.5 mag) and searching for companions on wide angular scales.  One other target (HD 31647) was observed on UT October 21, 2021 with a 5 telescope setup (missing telescope W2) using both MIRC-X (R190) and MYSTIC (prism, R50) \citep{Monnier2018SPIE10701E..22M}, a beam combiner in the same vein as MIRC-X which operates in the K-band ($\sim$ 2.2 microns), making our sample size a total of 27.  

For each target, we had the same observing sequence: 10 min source integration, the standard shutters sequence \citep{Anugu2020}, and repeated 10 min source integration.  For HD 56537, we performed 5 min source integrations instead due to the high flux of the primary to achieve equivalent S/N to all other targets.  Throughout our run, we observed calibrator stars for every few targets using an observing sequence of 5-10 min integrations, shutters, followed by 5-10 min integrations.  Two of our calibrators (HD 15734 and HD 78234) are resolved binaries, and instead we opted to use science targets well fit by a single star model to calibrate other targets observed in proximity, see Table \ref{table2}.  We opted to observe more science targets instead of more frequent calibrators because our search for companions relies on closure phases and does not need highly calibrated visibilities, see Sec. \ref{subsec:dataanalysis}.

%Does this last sentence make sense?

\begin{deluxetable}{cccccc}
\tablenum{1}
\tablecolumns{6} 
\tablecaption{Table of sources in our sample with Modified Julian Date (MJD) of observation.  Listed spectral types, ages, and masses for each star were taken from \citet{DeRosa2014} who describe their method of estimating age and mass in their Appendix.  Distances and their errors (16\% and 84\% confidence level) were extracted from the Gaia DR3 archive \citep{TheGaiaMission2016,Gaia_Multiplicity_2022arXiv220605595G, Babusiaux2022arXiv220605989B}, except where noted.}
\tablehead{\colhead{Target Name} &\colhead{SpType} &\colhead{Distance} &\colhead{Age}  &\colhead{Mass}   &\colhead{MJD} \\ \colhead{} &\colhead{} &\colhead{(pc)} &\colhead{(Myr)}  &\colhead{(M$_{\odot}$)}   &\colhead{}}
\startdata
HD 5448 & A5V & 42.37$^{+0.11}_{-0.19}$ & 450 & 2.39 & 59203.08\\
HD 11636$^{a}$  & A5V & 18.27$^{+0.25}_{-0.25}$ & 630 & 2.01 & 59204.08\\
HD 15550  & A9V& 71.77$^{+1.84}_{-0.56}$ & 790 & 1.84 & 59204.14\\
HD 20677  & A3V& 48.11$^{+0.30}_{-0.25}$ & 250 & 2.11 & 59203.18\\
HD 21912  & A3V& 56.29$^{+0.16}_{-0.15}$ & 40 & 1.77 & 59203.16\\
HD 24809  & A8V& 63.76$^{+0.12}_{-0.11}$ & 100 & 1.7 & 59203.22\\
HD 28910  & A8V& 46.97$^{+0.38}_{-0.37}$ & 630 & 2.21 & 59204.22\\
HD 29388$^{a}$  & A6V& 47.1$^{+1.2}_{-1.2}$ & 630& 2.17 & 59204.19\\
HD 31647*  & A1V& 49.91$^{+0.29}_{-0.29}$ & 30 & 2.39 & 59508.56\\
HD 32301  & A7V& 57.55$^{+4.80}_{-1.87}$ &630 & 2.22 & 59204.25\\
HD 46089$^{a}$ & A3V& 63.7$^{+1.5}_{-1.5}$ & 560 & 2.20 & 59203.29\\
HD 48097$^{a}$  & A2V& 43.6$^{+1.3}_{-1.3}$ & 30 & 1.94 & 59203.32\\
HD 56537$^{a}$  & A3V& 30.9$^{+0.2}_{-0.2}$ &320 & 2.39 & 59204.28\\
HD 59037  & A4V& 55.91$^{+3.84}_{-1.44}$ &500 & 2.16 & 59203.35\\
HD 66664  & A1V & 65.89$^{+0.70}_{-0.58}$ &320 & 2.42 & 59204.33\\
HD 74198$^{a}$  & A1IV & 55.6$^{+0.6}_{-0.6}$ &320 & 2.49 & 59204.39\\
HD 74873   &  A1V & 54.82$^{+0.15}_{-0.14}$ &50 & 1.88 & 59204.36\\
HD 77660  & A8V & 78.28$^{+0.20}_{-0.19}$ &710 & 1.81 & 59203.43\\
HD 84107  & A2IV & 51.24$^{+2.38}_{-0.98}$ & 10& 1.44 & 59203.45\\
HD 92941$^{a}$  &  A5V & 66.9$^{+1.4}_{-1.4}$ &450 & 1.84 & 59204.48\\
HD 97244  & A5V & 62.19$^{+0.18}_{-0.15}$ &60 & 1.72 & 59204.45\\
HD 99787  & A2V & 69.64$^{+0.68}_{-0.89}$ & 280& 2.32 & 59203.48\\
HD 106661  & A3V & 66.01$^{+0.09}_{-0.21}$ & 400& 2.29 & 59204.51\\
HD 112734  & A5 & 73.73$^{+0.44}_{-0.30}$ & 40& 1.69 & 59203.53\\
HD 115271$^{a}$  & A7V & 74.1$^{+2.4}_{-2.4}$ & 560 & 2.10 & 59203.59\\
HD 120047  & A5V & 52.84$^{+0.79}_{-0.39}$ & 500 & 1.78 & 59204.60\\
HD 121164 & A7V & 73.60$^{+0.58}_{-0.44}$ & 500 & 1.97 & 59204.57
\enddata
\tablenotetext{*}{Observed with both MIRC-X and MYSTIC at CHARA Array.}
\tablenotetext{a}{Distance taken from \citet{DeRosa2014} using the Hipparcos catalog \citep{Hipparcos1997ESASP1200.....E}}

\label{table1}
\end{deluxetable}

\begin{deluxetable}{cccc}
\tablenum{2}
\tablecolumns{4} 
\tablecaption{Table of calibrators.}
\tablehead{\colhead{Calibrator} &\colhead{UD diameter} &\colhead{Night}  &\colhead{UD Reference} \\ \colhead{Name} &\colhead{(mas)} & \colhead{UT} & \colhead{} }
\startdata
HD 99787 & 0.303 $\pm$ 0.024 & 20 Dec. 2020 & 1 \\
HD 44851 & 0.58 $\pm$ 0.014 & 20 Dec. 2020 & 2 \\
HD 21912 & 0.276 $\pm$ 0.007 & 20 Dec. 2020 & 2 \\
HD 120047 & 0.308 $\pm$ 0.008 & 21 Dec. 2020 & 2 \\
HD 32301 & 0.479 $\pm$ 0.033 & 21 Dec. 2020 & 2 \\
HD 74198 & 0.362 $\pm$ 0.024 & 21 Dec. 2020 & 2 \\
HD 19066* & 0.85 $\pm$ 0.06 & 21 Oct. 2021 & 2
\enddata
\tablenotetext{*}{Observed with both MIRC-X and MYSTIC at CHARA Array.}
\tablenotetext{1}{\citet{Swihart2017AJ....153...16S}}
\tablenotetext{2}{\citet{Bourges2017yCat.2346....0B}}
\label{table2}
\end{deluxetable}

\subsection{Data Reduction} \label{subsec:datareduction}
We used the standard MIRC-X data pipeline (version 1.3.3) to produce OIFITS files for each night, described in \citet{Anugu2020}. The MIRC-X pipeline and its documentation is maintained on Gitlab \footnote{\url{https://gitlab.chara.gsu.edu/lebouquj/mircx_pipeline}, remember to include the underscore in copied link}. This pipeline measures the visibilities, closure phases, and differential phases from each baseline pair in the raw interferometric data. We reduced our data with most of the default reduction parameters, setting the number of coherent integration frames (ncoh) to 10. We used an oifits maximum integration time of 60 seconds, which is lower than the default value of 150 seconds. This allows us to search for wider binaries, which create signals in visibility and phase that vary on faster timescales.

To calibrate the data and produce our final OIFITS files we used a modified version of the MIRC-X pipeline, which uses similar routines in IDL as the pipeline for the previous Michigan InfraRed Combiner (MIRC, \citealt{Monnier2007Sci...317..342M, zhao2009ApJ...701..209Z, che2011ApJ...732...68C, monnier2012}). This routine is well-tested with previous MIRC data to properly flag and remove bad data, which can otherwise corrupt binary fits and lead to worse non-detection maps.  The reduced oifits files of each source within our sample is available to download with this publication or on the CHARA data reduction machine\footnote{\url{https://www.chara.gsu.edu/observers/database}} which is open to the public upon request of an account.

%\textcolor{red}{Tyler: I should probably list all of the calibrators we used for each night, and the UDs assumed.}
\bigskip

\begin{figure*}[htb]
\gridline{\fig{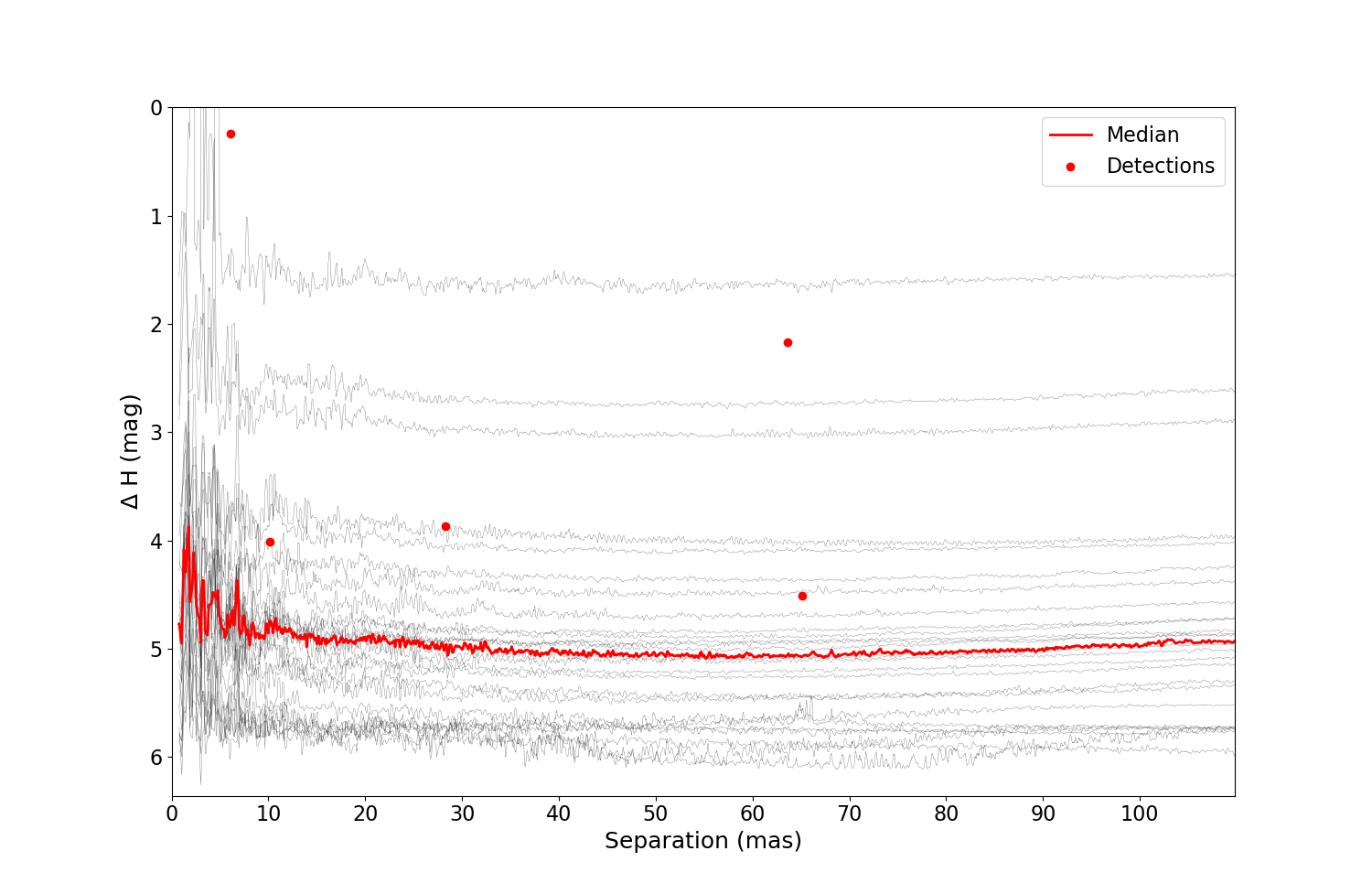}{\textwidth}{}}
\caption{Detection limits derived from CANDID for each source are plotted in black with the median value at each separation displayed in red.  Detection limits correspond to the highest companion flux at a given separation, over all position angles, that results in a 5$\sigma$ detection. Red circles represent the detections made in our survey at their given separation and contrast.}
\label{fig:binaryimages}
\end{figure*}

\subsection{Data Analysis} \label{subsec:dataanalysis}
In order to identify companions to the targets in our sample of A-type stars, we applied the open source python code CANDID \citep[Companion Analysis and Non-Detection in Interferometric Data,][]{Gallenne2015} to the OIFITS data of each observation.  This code is designed to identify companions through a grid search of three dimensional space (position and flux ratio).  It can use information from the visibilities, bi-spectrum amplitudes, and/or closure phases to perform binary model fitting for a given dataset. Closure phase is an interferometric quantity derived from the fringe phases of each pair of apertures within a closed triangle of three baselines (e.g. three telescopes), utilized to remove phase errors  \citep{Jennison1958MNRAS.118..276J,Rogers1974ApJ...193..293R, Monnier2000plbs.conf..203M, Monnier2004ApJ...602L..57M}.  For our purposes, we only used the closure phase information to find companions.

A symmetric brightness distribution will have measured closure phases of 0 or 180 degrees within the errors of the measurements for each combination of three telescopes.  Any scene deviating from a point-symmetric object (e.g. a binary) will produce non-zero closure phases with amplitudes on the order of the amount of deviation (e.g. flux ratio).  In this case, the grid search of CANDID will find the best-fit binary model that best replicates the observed closure phases including those deviating from zero.

CANDID provides a figure of merit in terms of n$\sigma$ calculated from $\chi^{2}$ statistics, defined in \citet{Gallenne2015}, to differentiate between a single source (uniform disk) and a binary model from the searched grid space.  First, we used the closure phase and visibility information to best estimate the diameter of the primary star with CANDID, which we then used as the diameter of the uniform disk model.  We then searched for companions within the closure phase data out to separations of 0.3", the extent of the interferometric field of view in grism mode.  We define 5$\sigma$ as the cutoff for companion detection as suggested by the CANDID authors.

Upon detection of a companion, we re-run CANDID using all available variables to derive the best fit binary model to the closure phase and visibility data.  CANDID allows for the fitting of the diameter of both components of the binary as well as fitting the "resolved flux" value which represents the visibility at a baseline of zero, typically accounting for any errors in calibration.  For binaries where we can resolve both components, our final binary model has six parameters: separation and position angle of the companion relative to the primary, the diameters of each components, the flux ratio between components, and the resolved flux of the visibility data.  For binaries with only a resolvable primary, we assume a diameter of 0.01 mas for the companion and use a five parameter binary model. See Table 3 showing the values of each of these parameters for our detections.

\begin{deluxetable*}{ccccccccc}
\tablenum{3}
\tablecolumns{9} 
\tablecaption{Detected binaries with resolved flux and flux ratio in percent relative to the primary, projected separations in arcseconds, position angles in degrees, uniform disk diameter of the primary and secondary in milli-arcseconds, and the reduced chi squared test statistic of a single star model and binary star model.  All parameters were derived using CANDID fitting to the closure phase and visibility data.  All detections achieved the maximum significance threshold on CANDID, 8$\sigma$.}
\tablehead{
\colhead{Target} & \colhead{Resolved Flux} & \colhead{Flux Ratio} & \colhead{Projected} & \colhead{PA (deg)} & \colhead{UD$_{1}$} &\colhead{UD$_{2}$} & \colhead{$\chi^{2}_{\nu, 1}$} & \colhead{$\chi^{2}_{\nu,2}$} \\
\colhead{Name} & \colhead{($\%$ Primary)} & \colhead{($\%$ Primary)} & \colhead{Sep. (mas) } & \colhead{(E of N)} & \colhead{(mas)} & \colhead{(mas)}& \colhead{}& \colhead{}}
\decimalcolnumbers
\startdata
  HD 5448 & -3.61$^{+0.64}_{-0.70}$ & 1.573$^{+0.039}_{-0.042}$ & 65.132 $\pm$ 0.007 & 144.324 $\pm$ 0.007 & 0.695 $\pm$ 0.002 & - & 1.75 & 1.22 \\
  HD 11636 & 1.97$^{+0.34}_{-0.38}$ & 13.546$^{+0.11}_{-0.089}$ &  63.632 $\pm$ 0.002 &  102.271 $\pm$ 0.002 & 1.0819 $\pm$ 0.0008 & 0.549 $\pm$ 0.007 & 159.5 & 3.18\\
  HD 28910 & 4.09$^{+0.72}_{-0.87}$ & 80.146$^{+0.064}_{-0.067}$ &  6.135 $\pm$ 0.0013 &  304.81 $\pm$ 0.012 & 0.377$^{+0.007}_{-0.006}$ & 0.341$^{+0.008}_{-0.007}$ & 1332 & 1.17\\
  HD 29388 & 1.535$^{+0.29}_{-0.32}$ & 2.488$^{+0.024}_{-0.023}$ &  10.180 $\pm$ 0.002 &  23.59 $\pm$ 0.02 & 0.553 $\pm$ 0.005 & - & 4.39 & 1.14\\
  HD 48097 & 7.25$^{+0.69}_{-0.66}$ & 2.84$^{+0.119}_{-0.11}$ &  28.296 $\pm$ 0.007  & 14.16 $\pm$ 0.02 & 0.288 $\pm$ 0.015 & - & 2.26 & 1.51
\enddata
\label{table3}
\end{deluxetable*}

\begin{deluxetable}{ccccc}
\tablenum{4}
\tablecaption{Physical separation in au, masses (\(\textup{M}_\odot\)), and mass ratios (q).  Companion masses are estimated using the MIST evolutionary models and the assumed primary mass and age from \citet{DeRosa2014}, see Table \ref{table1}.}
\tablewidth{0pt}
\tablehead{
\colhead{Target}& \colhead{Physical} & \colhead{M$_{prim}$} & \colhead{M$_{sec}$} & \colhead{q}\\
\colhead{Name} & \colhead{Sep. (au)} & \colhead{(M$_{\odot}$)} & \colhead{(M$_{\odot}$)} & \colhead{} }
\decimalcolnumbers
\startdata
  HD 5448 & 2.760 $\pm$ 0.013 & 2.39 &  0.60 & 0.25 \\
  HD 11636 &  1.145 $\pm$ 0.013 & 2.01 & 1.05 & 0.52\\
  HD 28910* &  0.288 $\pm$ 0.002 & 2.21 & 2.12 & 0.96\\
  HD 29388 &  0.480 $\pm$ 0.012 & 2.17 &  0.67 & 0.31\\
  HD 48097 &  1.23 $\pm$ 0.04  & 1.94 & 0.40 & 0.21\\
\enddata
\tablenotetext{*}{Mass may be overestimated due to high mass ratio and prior method of estimation. Mass ratios are still reliable. Cf. other A8V stars in sample.}
\label{table4}
\end{deluxetable}

\section{Results} \label{sec:results}

\subsection{Detections} \label{subsec:detections}
We detected five companions out of the 27 A-type stars in our sample.  All detections had a significance of 8$\sigma$, the maximum permitted value in CANDID, indicative of strong detections.  These companions were detected at projected separations = 6-65 mas (physical separation = 0.288 - 2.760 au) with flux ratios = 1.5 - 80\%, see Fig. \ref{fig:binaryimages} and Table 3.  Two of these detections, HD 11636 and HD 28910, were previously reported as spectroscopic binaries  \citep{Abt1965, Pourbaix2000A&AS..145..215P} with only HD 11636 having reliable orbital parameters.  Based on the inclination, eccentricity, and semi-major axis estimates of \citet{Pourbaix2000A&AS..145..215P}, we estimate the range of possible projected separations for HD 11636 to be from 4-68 mas, consistent with our detection at 64 mas near apocenter.  All five targets to which we detected a companion (except HD 11636, too bright for reliable Gaia information) were reported as having a proper motion anomaly using Hipparcos and Gaia catalogs that could be indicative of a companion \citep{Kervella2019A&A...623A..72K}.  However, only one of the five detections we made (HD 28910) was identified in Gaia Data Release 3 (DR3) as a non-single star \citep{Gaia_Multiplicity_2022arXiv220605595G}.  Gaia DR3 reports a period of 58.94 $\pm$ 0.09 days and an eccentricity of 0.31 $\pm$ 0.16.  Given our mass estimate of this system, see Table 4, and the period from Gaia, we estimate a semi-major axis of 0.48 au and a minimum and maximum distance from the primary as 0.33 and 0.63 au assuming the eccentricity from Gaia.  Our estimate of 0.288 au is within the errors of the current Gaia estimate, indicative of a larger eccentricity than the median Gaia value. Only one other source in our sample was identified in Gaia DR3 as being a non-single star, HD 21912, which has a spectroscopically identified companion with a period of 0.92 days \citep{Gaia_Multiplicity_2022arXiv220605595G}, likely at a separation smaller than our sensitivity.  All other sources within our sample had no companion detections in either our analysis or Gaia DR3.

Because the estimated ages of our targets range from 10-790 Myr, see Table \ref{table1}, we estimated the masses of each companion using the MIST evolutionary models \citep{Paxton2011ApJS..192....3P, Paxton2013ApJS..208....4P, Paxton2015ApJS..220...15P,Paxton2018ApJS..234...34P, Choi2016ApJ...823..102C, Dotter2016ApJS..222....8D}.  We assumed the primary mass and age estimates from \citet{DeRosa2014}.  They estimated ages based on the source position on the color-magnitude diagram relative to theoretical isochrones, and estimated masses based on their 2MASS K-band magnitude and estimated ages.  We adopt these values for the primary stars and use the MIST isochrones to determine companion masses from the magnitude differences derived in CANDID.  The mass ratios (q) of the detected multiple systems range from 0.21 - 0.96. The detections and the derived quantities of each system are listed in Table 4.

\subsection{Detection Limits}\label{subsec:detectionlimits}

CANDID offers a feature that derives the detection limit to any particular target.  This is accomplished by injecting artificial companions at particular positions with specific flux ratios and evaluating the significance of a binary fit with those given values.  The code then identifies the flux ratio at each step in separation where the significance level is equal to 5$\sigma$.  Because it searches the entire field of view, any given separation would have many injected companions at many position angles.  To determine the detection limit at any given separation without position angle bias, we take the highest flux ratio that returns the 5$\sigma$ result.  In this case, there could be fainter companions at other position angles that correspond to the 5$\sigma$ limit.  This process results in a flux ratio at all steps in separation in which we can recover companions.  For targets with detected companions, CANDID also offers the ability to remove the signal of the detection and then carry out the detection limit procedure on the edited data.  For sources with detected companions, we defined companion parameters by fitting the closure phase and visibility data for all free parameters, removed the signal, and then performed the detection limits procedure on the residual closure phase data.

Achievable contrasts typically range from $\Delta$H = 4.0-5.5 mag beyond $\sim$ 10 mas.  The H-band mag of our targets range from 2.43 - 6.39 mag.  For targets with H $\lesssim$ 5.8 mag, we can achieve typical contrasts under good conditions. For target with H $\gtrsim$ 5.8 mag, the achievable contrast is $<$  4 mag at 10 mas, especially with worse weather conditions.  In Fig. \ref{fig:binaryimages}, we show the detection limits for each source in our sample with the median value at each separation highlighted in red and the values of the detections represented by red circles.  In Table \ref{sensitivity}, the attainable contrast at select angular separations is given for each source within the sample.

Using the achievable contrast at a given separation, we can calculate the lowest mass of a detectable companion for each target as a function of separation.  Given the age (ranging from 10-790 Myr) and mass of the primary from \citet{DeRosa2014}, we use the MIST evolutionary models to estimate the lowest mass of a detectable companion, and the lowest mass ratio of a detectable multiple system, at each separation.

Four targets stand out as having poor sensitivity (higher mass ratio limits) relative to the rest of the sample.  All four were observed on UT Dec. 20, 2020 the night with poorer observing conditions.  Three of the four are the faintest sources in our sample and have H-band $\geq$ 5.9 mag (HD 77660, HD 24809, and HD 112734).  The other source (HD 115271) was observed at the end of the first night during sunrise, resulting in poor quality data.

\begin{figure*}[htb]
\gridline{\fig{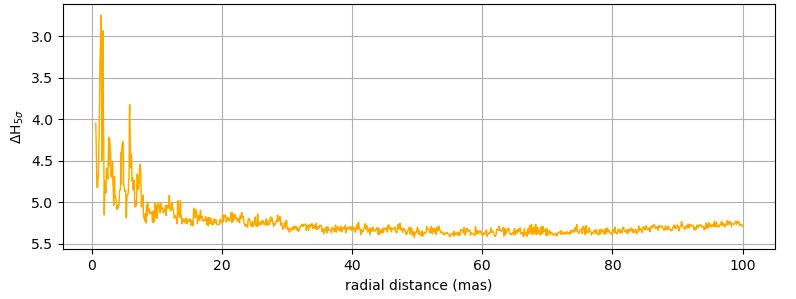}{\textwidth}{MIRC-X}}
\gridline{\fig{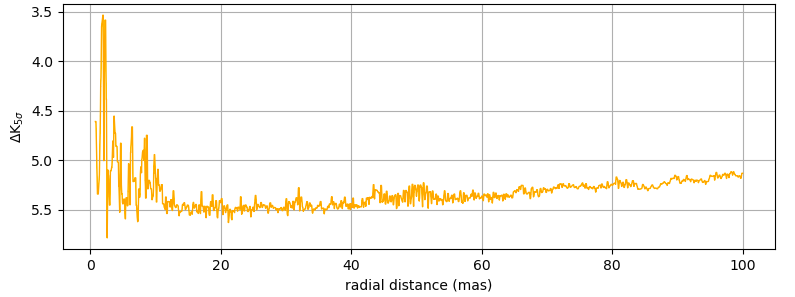}{\textwidth}{MYSTIC}}
\caption{Derived detection limit for HD 31647 as a function of radius from the central star on both the MIRC-X and MYSTIC instrument.  These data were collected simultaneously with 15 min integrations.}
\label{fig:omeaur}
\end{figure*}

One target (HD 31647) was observed with both MIRC-X and MYSTIC using 5 telescopes (instead of all 6) with 15 min total integration time.  HD 31647 has H = 4.98 mag, a typical brightness for our sample, and we are able to detect companions at the 5$\sigma$ level down to contrast of 5 mag beyond 10 mas with both MIRC-X and MYSTIC.  The achievable contrast was similar between both instruments as shown in Fig. \ref{fig:omeaur}, but slightly better for MYSTIC.  

%\begin{figure*}[htb]
%\gridline{\fig{Ex_5sig_cc_chara.png}{\textwidth}{}}
%\caption{Derived detection limit for HD 74198 as a function of radius from the central star.}
%\label{fig:detectionlimitexample}
%\end{figure*}

%DETECTION LIMIT TABLE
\begin{deluxetable}{c|cccccc}
\tablenum{5}
\tablecolumns{7}
\tablecaption{Detection limits (contrast in units of magnitudes for the H band) derived for each target using MIRC-X at 1.0, 3.0, 5.0, 10.0, 50.0, and 300.0 milli-arcseconds (mas) in angular separation. We define the detection limit as the highest flux ratio companion that CANDID can recover at a given radius that is equivalent to a 5$\sigma$ detection.  The binary detections are labeled below where the listed detection limits are derived after the removal of the companion with the subsequent injection and recovery procedure. \label{sensitivity}}
\tablewidth{0pt}
\tablehead{\colhead{} & \colhead{[mas]} \\
\colhead{Target Name} & \colhead{1.0} & \colhead{3.0}& \colhead{5.0}& \colhead{10.0}& \colhead{50.0}& \colhead{300.0}}
\startdata
HD 5448$^{a}$ & 4.80 & 4.80 & 5.01 & 5.49 & 5.71 & 5.06\\
HD 11636$^{a}$ & 5.27 & 5.39 & 5.40 & 5.67 & 5.67 & 4.02 \\
HD 15550  & 3.74 & 4.08 & 4.22 & 4.43 & 4.82 & 4.19 \\
HD 20677  & 3.97 & 3.98 & 4.25 & 4.77 & 5.03 & 4.44 \\
HD 21912  & 3.72 & 3.75 & 4.14 & 4.56 & 4.88 & 4.05 \\
HD 24809  & 0.98 & 1.54 & 1.99 & 2.37 & 2.71 & 2.02 \\
HD 28910$^{a}$ & 3.73 & 4.10 & 4.20 & 4.48 & 5.09 & 4.28 \\
HD 29388$^{a}$ & 4.14 & 5.15 & 5.47 & 5.49 & 5.70 & 5.34 \\
HD 31647 & 3.44 & 4.55 & 4.59 & 5.04 & 5.40 & 4.80 \\
HD 32301  & 5.03 & 5.03 & 5.22 & 5.55 & 5.94 & 5.48\\
HD 46089 & 2.72 & 3.32 & 3.66 & 3.93 & 4.46 & 3.90 \\
HD 48097$^{a}$ & 2.91 & 3.10 & 3.61 & 3.66 & 4.10 & 3.48 \\
HD 56537  & 4.82 & 5.20 & 5.44 & 5.50 & 5.82 & 5.50 \\
HD 59037  & 3.92 & 4.16 & 4.37 & 4.66 & 4.93 & 4.36\\
HD 66664  & 3.39 & 4.53 & 4.74 & 4.99 & 5.43 & 4.88\\
HD 74198   & 4.87 & 4.95 & 5.38 & 5.63 & 5.83 & 5.46 \\
HD 74873   & 3.00 & 4.08 & 4.23 & 4.62 & 5.06 & 4.47 \\
HD 77660  & 1.33 & 1.76 & 2.28 &2.60 & 2.99 & 2.37 \\
HD 84107  & 3.15 & 3.15 & 3.48 & 4.02 & 4.33 & 3.73 \\
HD 92941  & 3.80 & 4.12 & 4.39 & 4.58 & 4.92 & 4.26 \\
HD 97244  & 3.66 & 3.66 & 3.92 & 4.28 & 4.67 & 4.05 \\
HD 99787  & 3.87 & 3.97 & 4.16 & 4.65 & 4.98 & 4.19 \\
HD 106661  & 4.74 & 4.83 & 5.01 & 5.26 & 5.59 & 5.03 \\
HD 112734  & 0.54 & 0.54 & 1.01 & 1.26 & 1.61 & 0.81 \\
HD 115271  & 2.90 & 3.03 & 3.29 & 3.39 & 3.96 & 3.46 \\
HD 120047  & 4.00 & 4.00 & 4.23 & 4.96 & 5.23 & 4.58 \\
HD 121164 & 3.99 & 3.99 & 4.37 & 4.69 & 5.20 & 4.56 \\
\enddata
\tablenotetext{a}{Binary detections. Limits derived after removal of companion from data.}
\label{table5}
\end{deluxetable}

\subsection{Binary Population Analysis} \label{subsec:population}

Out of 27 A-type stars observed at CHARA, we detected five companions.  We must take into account our sensitivity to companions for each source in terms of flux ratio and separation in order to adequately characterize this stellar population.  

We first identify a region of common sensitivity in terms of projected separation and mass ratio for the vast majority of our sample that incorporates as many of our detections as possible.  We can detect companions down to mass ratios of 0.25 at projected separations of 0.288 au for 20 of the 27 targets in our sample.  One target (HD 28910) has a bright companion detection where the signal removal process in CANDID cannot adequately remove all companion features.  The achievable mass ratio at the location of the companion is underestimated, while those at locations away from the companion reach the mass ratio of common sensitivity.  Therefore, we include this one as having sufficient sensitivity due to both its S/N relative to the rest of the sample and reaching the mass ratio of 0.25 at closer and wider separations.  This results in 21 out of 27 targets sensitive to mass ratios $\geq$ 0.25 at separations $\geq$ 0.288 au.

We must also define the common outer working angle of our sample due to the difference in distances as defined in the Gaia archive \citep{TheGaiaMission2016, earlygaia2021A&A...649A...1G, Babusiaux2022arXiv220605989B}.  For the sake of retaining as many detections as possible and including as many targets as possible in this sub-sample, we set the outer boundary of our common sensitivity to 5.481 au, the largest detectable companion separation for HD 11636 due to its proximity to the Sun (18.27 pc).  Therefore, we require the sensitivity to companions down to mass ratios of 0.25 at separations of 0.288 - 5.481 au in order to be included in this sub-sample.  We find that for these 21 A-type stars (M = 1.44 - 2.49 M$_{\odot}$) in our sample there are four detections, resulting in a companion frequency of 0.19$^{+0.11}_{-0.06}$ over mass ratios 0.25 - 1.0 and projected separations 0.288 - 5.481 au.  Error bars are derived using the binomial distribution formalism of \citet{Burg2003} appropriate for our small sample.

\section{Discussion} \label{sec:discussion}

Our source list was derived from the \citet{DeRosa2014} VAST sample.  We excluded sources with known Ap or Am spectral types, but did not make sample selections based on previously detected companions.  Sources were then randomly selected based on the nights allocated.  Of the five binary detections we made over separations of 0.288-2.760 au, three of the sources were targeted for companion search through adaptive optics imaging and common proper motion analysis.  They did not find companions to those three over separations sampled, and they are not known to be triple systems.  HD 5448 was observed from 32 - 794 au and 3980 - 45000 au.  HD 11636 was observed from 32 - 158 au and 3980 - 45000 au.  HD 48097 was observed from 70 - 354 au and 1780 - 45000 au.

%\textcolor{red}{Gaia is sensitive enough to observe all of our targets.  However, the angular resolution of Gaia is not good enough to resolve any of our detected companions, $\sim$ 0.4" \citep{earlygaia2021A&A...649A...1G, earlygaia2021A&A...649A...1G}.  \citet{Fabricius2021A&A...649A...5F} show that the completeness of recovering known visual binaries drops below 90\% around 1” and is nearly 0\% at 0.4” for the early Data Release 3.  Gaia also takes spectra of all of the observed targets, attempting to derive radial velocities to all sources.  However, for sources with T$_{eff}$ $>$ 7000 K, this can be difficult.  The Paschen lines are blended with the Ca II lines which are highly dependent on surface gravity and can shift the centroid of the spectral feature, increasing the uncertainty in the radial velocity measurement by several km/s \citep{katz2019A&A...622A.205K}.  Therefore, it is difficult to predict the ability of Gaia to detect companions to A-type stars through the radial velocity method.  The best opportunity for Gaia to detect these companions is through astrometric motion which is expected to detect companions with periods of 0.03 – 30 years in DR4\footnote{\url{https://sci.esa.int/web/gaia/-/31441-binary-stars}}.  The astrometric information will have fitted mass functions for each detected multiple and will likely be able to detect these companions, although not much information is available to determine what combination of separations, mass ratios, and eccentricities will be detectable through astrometric monitoring.}

\subsection{Comparing the Mass Ratios to Models} \label{subsec:massratiocomparison}

We cannot place meaningful constraints on the mass ratio distribution of our own sample, because we only have five detections and model fitting suffers from small number statistics.  Interestingly, \citet{Moe2017} and \citet{ElBadry2019MNRAS.489.5822E} both describe a model of the companion population to A-stars as weighted towards smaller mass ratios with an excess twin binary component, analyzing results for wider companions from \citet{DeRosa2014}.  Although we only have five detections, four of those five are smaller mass ratios between q=0.21-0.52 where one (q=0.96) is considered a twin (q $\geq$ 0.95).

Even with a small number of detections, we can compare models of the mass ratio distribution to our sample through the Kolmogorov–Smirnov (KS) test.  The \citet{DeRosa2014} multiplicity survey of A-type stars is sensitive to companions beyond $\sim$ 30 au and mass ratios $\geq$ 0.15 for a majority of their sample, and describe the mass ratio distribution as a power law for q $\geq$ 0.15:

\begin{equation}
    \frac{dN_{1}}{dq} \propto q^{\beta}
    \label{qdistribution}
\end{equation}

For companions between 30-125 au and 125 - 800 au, the best-fit to the mass ratio distribution has a power-law index $\beta$ = -0.5$^{+1.2}_{-1.0}$ and -2.3$^{+1.0}_{-0.9}$ respectively.  Over mass ratios 0.25-1.0, we cannot reject the null hypothesis that our four detections (excluding one with q=0.21 due to being outside the sensitivity of the sub-sample) are drawn from the same parent population as the distribution of inner binaries ($\beta$ = -0.5) or outer binaries ($\beta$ = -2.3) from \citet{DeRosa2014} with a KS-test p-value = 0.48 and 0.77 respectively, where a larger sample size is needed to make such a distinction.

\subsection{Companion Frequency Comparison} \label{subsec:surveycomparison}
Previous surveys have estimated the companion frequency to A-type stars with methods other than long-baseline interferometry.  \citet{Murphy2018MNRAS.474.4322M} used a phase modulation technique on Kepler light curves \citep[see][]{Murphy2014MNRAS.441.2515M} to identify companions to $\delta$ Scuti variable stars (type A/F stars).  With these data, they are sensitive to companions with periods $\gtrsim$ 100 days where incompleteness becomes large, and $<$ 1500 days, or a $\sim$ 0.6 - 3.6 au.  They found a companion frequency of 0.139 $\pm$ 0.021 after accounting for white dwarf companions (evolved higher mass stars that would have served as the primary before evolution).  Roughly two-thirds of their companions have q $>$ 0.25 after accounting for incompleteness, for a companion frequency of $\sim$ 0.095 $\pm$ 0.021.  For our sample over a = 0.6 - 3.6 au \citep[the sensitivity of][]{Murphy2018MNRAS.474.4322M} and q $\geq$ 0.25, we detected two companions out of the 21 sources in our sub-sample for a companion frequency of 0.1$^{+0.1}_{-0.03}$, consistent with their results.

\citet{Abt1965} searched for companions to A-type stars using radial velocity measurements from many different publications.  It is difficult to constrain the sensitivity of this program (i.e. mass ratio limits) due to undefined errors in some of the measured velocities from past observations.  However, they give a measurement error of $\sim$ 2 km/s for their own measurements, meaning they should be sensitive to stellar companions to intermediate mass stars from periods of roughly 100-1500 days, the sensitivity of \citet{Murphy2018MNRAS.474.4322M}.  Nine of their detections have measured periods between 100-1500 days for a companion frequency of 0.16$^{+0.06}_{-0.04}$ or 0.13$^{+0.06}_{-0.04}$ after accounting for expected white dwarf companions (21$\%$ of expected companions).  These results are consistent with both the \citet{Murphy2018MNRAS.474.4322M} results and our own.

Additionally, our survey is sensitive over a defined range of orbital separations and mass ratios that may not correspond to the same sensitivity of previous surveys.  However, many large surveys have modeled the distribution of orbital separations and mass ratios of their companion population using functional forms.  We can take those functions, and integrate them over the sensitivity of our survey to derive an expected companion frequency (CF) of the model over our sensitivity.  Then, we can derive a probability that our observations can be described by this model, and determine whether the model is representative of our results.  We use the models for the companion population to solar-type primaries \citep{Raghavan2010}, A-type primaries \citep{DeRosa2014}, and B-type primaries \citep{Rizzuto2013MNRAS.436.1694R} to determine if any may be consistent with our results.

The mass ratio distribution is well-fit with a power law (see eq. \ref{qdistribution}) and the orbital separation distribution is well-fit with a log-normal distribution with a peak (log(a$_{o}$)) and a width ($\sigma_{loga}$):

\begin{equation}
\frac{dN_{2}}{da} = \frac{1}{\sqrt{2\sigma_{loga}^{2}}} e^{-\frac{(log(a)-log(a_{o}))^{2}}{2\sigma_{loga}^{2}}}
    \label{adistribution}
\end{equation}

Integrating the mass ratio distribution of eq. \ref{qdistribution} and the orbital separation distribution of eq. \ref{adistribution}, we can calculate the expected CF over specified separations and mass ratios  assuming they are independent, as shown in \citet{DeFurio2019, DeFurio2022ApJ...925..112D}:

\begin{equation}
    CF = C_{n}*\int_{q_1}^{q_2} \frac{dN_{1}}{dq} \int_{a_1}^{a_2} \frac{dN_{2}}{da}
    \label{cf}
\end{equation}

\begin{deluxetable*}{ccccccccc}
\tablenum{6}
\tablecolumns{9} 
\tablecaption{Companion population parameters and sensitivities of each tested model with the resulting statistics.}
\tablehead{\colhead{Primary} &\colhead{$\beta$} &\colhead{log(a$_{o}$)}  &\colhead{$\sigma_{loga}$} & \colhead{CF}& \colhead{log(a)}& \colhead{q} &\colhead{Expected CF} &  Posterior \\ \colhead{Spectral Type} & \colhead{}&\colhead{}&\colhead{}&\colhead{}&\colhead{sensitivity}&\colhead{sensitivity} &\colhead{} &\colhead{Probability} }
\startdata
FGK-type$^{1}$ & 0 & 1.7 & 1.68 & 0.58 $\pm$ 0.02 & -2 $\leq$ log(a) $\leq$ 4 & q $\geq$ 0.1 & 0.108 $\pm$ 0.002 & 0.08 \\
A-type$^{2}$ & -0.5$^{+1.2}_{-1.0}$ & 2.59 $\pm$ 0.13 & 0.79 $\pm$ 0.12 & 0.219 $\pm$ 0.026 & 1.5 $\leq$ log(a) $\leq$ 2.9 & q $\geq$ 0.1 & 0.002$^{+0.004}_{-0.0015}$ & 10$^{-6}$\\
A-type$^{2}$ & -0.5$^{+1.2}_{-1.0}$ & 2.59 $\pm$ 0.13 & 0.79 $\pm$ 0.12 & 0.338 $\pm$ 0.026 & 1.5 $\leq$ log(a) $\leq$ 4.0 & q $\geq$ 0.1 & 0.002$^{+0.004}_{-0.0015}$ & 10$^{-6}$\\
B-type$^{3}$ & -0.46 $\pm$ 0.14 & 1.05 $\pm$ 0.2 & 1.35 $\pm$ 0.2 & 1.35 $\pm$ 0.20 &  -2 $\leq$ log(a) $\leq$ 4 & q $\geq$ 0.1 & 0.20$^{+0.06}_{-0.05}$ & 0.46
\enddata
\tablenotetext{1}{\citet{Raghavan2010}}
\tablenotetext{2}{\citet{DeRosa2014}}
\tablenotetext{3}{\citet{Rizzuto2013MNRAS.436.1694R}}
\label{table6}
\end{deluxetable*}

First, we solve for C$_{n}$ by sampling from the companion distribution parameter values of each survey (see Table \ref{table6}) 10$^{5}$ times, assuming Gaussian errors and integrating over the stated sensitivities. Then, we used the value of the constant to calculate the expected CF assuming those companion parameters over the sensitivity of our sample, q=0.25-1.0 and a=0.288-5.481 au. We integrate the binomial distribution of the results of our A-star survey, 4 detections out of 21 targets, from 0 to the CF of each sampling stated above. We then take the mean of these samples to arrive at a posterior probability that this model can describe our observations of A-stars over q=0.25-1.0 and a=0.288-5.481 au.  

\begin{figure}[ht!]
\gridline{\fig{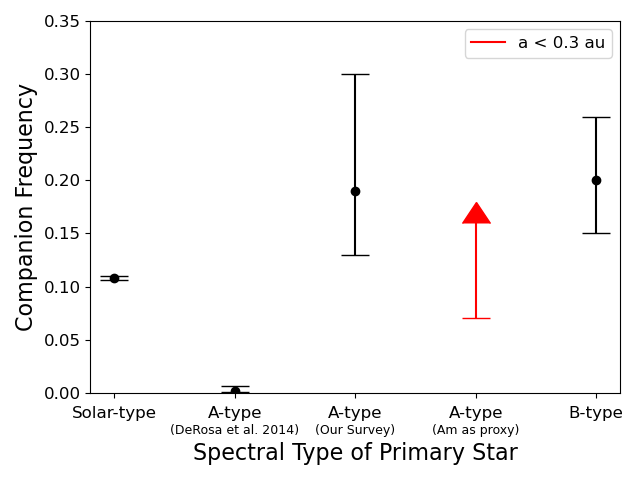}{0.45\textwidth}{}}
\caption{Companion frequencies based on spectral type of the primary star from the various models listed in Sec. \ref{subsec:surveycomparison}. All frequencies are calculated over mass ratios = 0.25 - 1.0 and separations (a) = 0.288 - 5.481 au, except for Am-stars which was calculated for a $<$ 0.3 au, see Sec. \ref{subsec:amcomparison}.  The companion frequency for Solar-type primaries is derived from the model of \citet{Raghavan2010}, A-type primaries from the model of \citet{DeRosa2014} and this survey, A-type primaries using Am stars from \citet{Carquillat2007}, and B-type primaries from the model of \citet{Rizzuto2013MNRAS.436.1694R}. }
\label{fig:cfs}
\end{figure}

The expected companion frequencies are summarized in Fig. \ref{fig:cfs} where they are presented according to the individual survey.  As shown in Table \ref{table6}, the estimated posterior probabilities for the companion population models to solar-type, A-type, and B-type primaries are 0.08, 10$^{-6}$, and 0.46 respectively.  These results allow us to conclude that: a) there is a significant population of close companions to A-type stars that must be modeled to have a complete picture of the companion population to intermediate mass stars, and b) we find no evidence for a difference between the multiple population of our A-type sample and that of either solar-type stars or B-type stars.  

It is important to note that \citet{Moe2017} refute the decrease of the separation distribution of \citet{DeRosa2014} for A-type primaries for a $<$ 100 au.  They argue that this decrease is due to incompleteness for q $<$ 0.2, and calculate a consistent companion frequency per decadal bin in the log of the orbital period across roughly a=20-50 au and a=50-450 au, consistent with flat in log-period space.  \citet{DeRosa2014} themselves argue that their model of the orbital separation distribution is inconsistent with the detected companions through radial velocity measurements of \citet{Abt1965}, and state that further observations are required to define the close companion population.

\subsection{Comparison to Radial Velocity Survey of Am Stars} \label{subsec:amcomparison}

\citet{AbtLevy1985} hypothesize that the Am star (chemically peculiar and slowly rotating) phenomena are the result of a close companion causing tidal braking which slows the rotation of normal A stars.  Once the rotation decreases below $\sim$ 100 km/s, diffusion within the star is allowed to bring metals to the surface which produces the metallic lines seen in Am stars \citep{Michaud1980AJ.....85..589M, Michaud1983ApJ...269..239M}.  However, not all Am stars appear to have close companions, and the presence of metallic lines may need to be explained without tidal braking.  One hypothesis is that as the star evolves off the main sequence, its expansion will slow down the rotation rate \citep{AbtLevy1985} below 100 km/s, allowing diffusion to proceed, producing the metallic lines seen in the Am spectra.  Other hypotheses such as accretion of interstellar material by distorted magnetic fields \citep{Bohm2006PASP..118..419B} could also explain the presence of higher metal abundances.

Estimates for the occurrence of Am stars among all A-type stars range from 15-35\% \citep{Bohm2006PASP..118..419B, Gray2016AJ....151...13G,  Hummerich2017MNRAS.466.1399H, Qin2019ApJS..242...13Q}.  Based on the results of \citet{Carquillat2007}, 47\% of their Am star sample have companions with periods $<$ 30 days ($\sim$ 0.3 au for an equal mass binary of 2 M$_{\odot}$ stars, comparable to our smallest separation detection).  If all A-type stars with very close companions become Am stars, then the binary frequency of Am stars at close separations can serve as a proxy for the binary frequency of all A-type stars at those same close separations.  Given the fraction of Am stars to all A-type stars (15-35\%) and the binary frequency of Am stars on short orbits \citep[47\%]{Carquillat2007}, we can expect that 7-16\% of all A-type stars have companions with periods $<$ 30 days ($\sim$ 0.3 au), see Fig. \ref{fig:cfs}.  This result is consistent with our observed companion frequency of 0.19$^{+0.11}_{-0.06}$ beyond 0.288 au, indicating that the binary frequency of A-type stars at separations $<$ 0.3 au may be comparable to that between 0.288 - 5.481 au.

\subsection{Implications} \label{subsec:implications}

Our analysis of 27 A-type stars within 80 pc observed with long-baseline interferometry reveals a companion frequency of 0.19$^{+0.11}_{-0.06}$ over mass ratios 0.25-1.0 and projected separations 0.288-5.481 au.  In comparison to other multiplicity studies of stellar populations, we find that an extrapolation of the fitted companion properties to A-type primaries of \citet{DeRosa2014} cannot reproduce the observed companion frequency of our A-star survey over our sensitivity.  However, they state that their model is inconsistent with previous detections at close separations and further investigations are required to define the companion population at small separations.  We could not rule out their best-fit mass ratio distribution (eq. \ref{qdistribution}) for either outer ($\beta$ = -2.3) or inner ($\beta$ = -0.5) binaries based on our results.  We do not find any strong statistical difference between the companion frequency of B-type stars in Sco-Cen \citep{Rizzuto2013MNRAS.436.1694R} and our sample of A-type stars, nor between that of solar-type stars in the field \citep{Raghavan2010} and our sample of A-type stars, over common mass ratios and separations sampled.  We also do not find any statistical differences between the very close ($<$ 0.3 au) companion frequency of Am stars and the companion frequency of normal A-type stars over our sensitivity.  These companion frequencies are summarized in Table \ref{table6} and Fig. \ref{fig:cfs} where they are presented according to the individual survey.

It is important to note that all five companions were found at separations between 0.288-2.760 au.  For many sources in our sample, our survey was sensitive to companions at separations greater than 2.760 au due to their distance (up to 78 pc) and our ability to detect companions out to 0.3", yet none were found.  It is a potentially curious observation, and a larger survey would allow us to probe whether this defecit of companions beyond 2.760 au is statistically significant.

Companions formed through turbulent fragmentation \citep{Goodwin2004, Offner2010} are thought to originate at separations of 100s-1000s au, while companions formed through disk fragmentation \citep{Adams1989, Bonnell1994, Kratter2008} are thought to originate at separations on the order of the size of the disk, 10-100s au.  Various processes can reduce the separation of the companion below 10 au \citep{Bate2003MNRAS.339..577B,Bate2012MNRAS.419.3115B}.  These processes include accretion from the natal cloud, interactions with the circumbinary disk and gas in the natal cloud, dynamical effects within an unstable triple, the Kozai-Lidov mechanism, and more \citep{ Offner2010,Kratter2011ASPC..447...47K, MoeKratter2018, Lee2019ApJ...887..232L, Tokovinin2020MNRAS.491.5158T}.  We cannot specifically determine which processes affected the binaries observed in our sample.  However, we can conclude that these processes are important to a significant portion of the A-star population and are required to replicate our observations. Although our results do not place strong constraints on the companion population to intermediate mass A-type stars, they indicate that such a population is significant and must be precisely characterized for a global understanding of star formation processes to intermediate mass stars.  A larger sample size is required which will enable us to: a) place strong constraints on the companion frequency to A-type stars at close separations, b) fit various mass ratio distributions for a larger set of companions, and c) probe the functional form of the close companion orbital separation distribution.

%\citet{DeRosa2014} found that the companion frequency of A-type stars is consistent with that of solar-type stars from \citet{Raghavan2010} over q $\geq$ 0.1 and a = 30 - 10,000 au.  We found an increase in companions relative to the \citet{Raghavan2010} survey over q $\geq$ 0.27 and a=0.3-3.3 au, 7\% vs. 19\% see Sec. \ref{subsec:surveycomparison}.  If A-type stars were more likely to retain companions due to turbulent fragmentation alone, we would expect a larger companion frequency at wider separations as well as closer separations relative to solar-type stars.  However, we see an increase in companion frequency only at small separations.  This may be the result of enhanced disk fragmentation for A-type stars relative to FGK-stars, consistent with model predictions that show an increase in disk fragmentation with stellar mass \citep{Kratter2006MNRAS.373.1563K, Kratter2008}.  Such a process could lead to a preferential accumulation of mass onto the companion as they migrate through and accrete from the circumstellar disk, increasing the mass ratio of the multiple system \citep{Kroupa1995MNRAS.277.1491K, Bate1997MNRAS.285...33B, Bate2000MNRAS.314...33B}.  

\section{Conclusion} \label{sec:conclusion}
We have conducted a survey of 27 nearby A-type stars within 80 pc using long baseline interferometry at the CHARA Array with the MIRC-X instrument to search for close companions.  To summarize the results of our survey:

1) We have shown that 20 min integrations with the Grism190 mode on targets with H $\leq$ 5.8 mag MIRC-X can achieve contrasts of 4-5.5 mag beyond $\sim$ 10 mas and 3-4.5 mag beyond 2 mas.  We have also shown that MYSTIC has comparable, if not slightly better, sensitivity to faint companions compared to MIRC-X.

2) We detected 5 companions with projected separations 6-65 mas (0.288-2.76 au) and mass ratios = 0.21-0.96.  For our sample of A-type stars with masses of 1.44-2.49 M$_{\odot}$, we observed a companion frequency of 0.19$^{+0.11}_{-0.06}$ over mass ratios 0.25-1.0 and projected separations 0.288-5.481 au.  This measurement is consistent with that of the spectroscopic survey of \citet{Abt1965} and the pulsation timing analysis of \citet{Murphy2018MNRAS.474.4322M} for AF-type primaries over separations and mass ratios sampled.

3) Our estimate of the companion frequency is larger than the calculated companion frequency after extrapolating the model fits of \citet{DeRosa2014} to the sensitivity of our survey with a probability of 10$^{-6}$ that they can describe our observations.  Although, they stated that further investigations are required to completely characterize the companion population down to small separations that are not included in their analysis.

4) Our detections are consistent with being drawn from both a flat and negatively weighted mass ratio distribution.  A larger sample size is crucial to characterizing the mass ratio distribution of close companions to A-type stars.

\begin{acknowledgments}
This work is based upon observations obtained with the Georgia State University Center for High Angular Resolution Astronomy Array at Mount Wilson Observatory. The CHARA Array is supported by the National Science Foundation under Grant No. AST-1636624 and AST-2034336. Institutional support has been provided from the GSU College of Arts and Sciences and the GSU Office of the Vice President for Research and Economic Development. Time at the CHARA Array was granted through the NOIRLab community access program (NOIRLab PropID: 2020B-0290; PI: M. De Furio). MIRC-X received funding from the European Research Council (ERC) under the European Union's Horizon 2020 research and innovation programme (Grant No. 639889). JDM acknowledges funding for the development of MIRC-X (NASA-XRP NNX16AD43G, NSF-AST 1909165) and MYSTIC (NSF-ATI 1506540, NSF-AST 1909165).  We would like to thank the Jean-Marie Mariotti Center for providing the tools ASPRO2 and SearchCal, used to plan observations, available at \url{http://www.jmmc.fr/aspro}.  S.K. acknowledges support from ERC Starting Grant and Consolidator Grant (Grant Agreement Nos.\ 639889 and 101003096).  This work has made use of data from the European Space Agency (ESA) mission
{\it Gaia} (\url{https://www.cosmos.esa.int/gaia}), processed by the {\it Gaia}
Data Processing and Analysis Consortium (DPAC,
\url{https://www.cosmos.esa.int/web/gaia/dpac/consortium}). Funding for the DPAC
has been provided by national institutions, in particular the institutions
participating in the {\it Gaia} Multilateral Agreement.
\end{acknowledgments}

\bibliography{bibliography}{}

\begin{thebibliography}{}
\expandafter\ifx\csname natexlab\endcsname\relax\def\natexlab#1{#1}\fi
\providecommand{\url}[1]{\href{#1}{#1}}
\providecommand{\dodoi}[1]{doi:~\href{http://doi.org/#1}{\nolinkurl{#1}}}
\providecommand{\doeprint}[1]{\href{http://ascl.net/#1}{\nolinkurl{http://ascl.net/#1}}}
\providecommand{\doarXiv}[1]{\href{https://arxiv.org/abs/#1}{\nolinkurl{https://arxiv.org/abs/#1}}}

\bibitem[{Hip(1997)}]{Hipparcos1997ESASP1200.....E}
 1997, ESA Special Publication, Vol. 1200, {The HIPPARCOS and TYCHO catalogues.
  Astrometric and photometric star catalogues derived from the ESA HIPPARCOS
  Space Astrometry Mission}

\bibitem[{{Abt}(1965)}]{Abt1965}
{Abt}, H.~A. 1965, \apjs, 11, 429, \dodoi{10.1086/190120}

\bibitem[{{Abt} \& {Levy}(1985)}]{AbtLevy1985}
{Abt}, H.~A., \& {Levy}, S.~G. 1985, \apjs, 59, 229, \dodoi{10.1086/191070}

\bibitem[{{Adams} {et~al.}(1989){Adams}, {Ruden}, \& {Shu}}]{Adams1989}
{Adams}, F.~C., {Ruden}, S.~P., \& {Shu}, F.~H. 1989, \apj, 347, 959,
  \dodoi{10.1086/168187}

\bibitem[{{Anugu} {et~al.}(2020){Anugu}, {Le Bouquin}, {Monnier}, {Kraus},
  {Setterholm}, {Labdon}, {Davies}, {Lanthermann}, {Gardner}, {Ennis},
  {Johnson}, {Ten Brummelaar}, {Schaefer}, \& {Sturmann}}]{Anugu2020}
{Anugu}, N., {Le Bouquin}, J.-B., {Monnier}, J.~D., {et~al.} 2020, \aj, 160,
  158, \dodoi{10.3847/1538-3881/aba957}

\bibitem[{{Babusiaux} {et~al.}(2022){Babusiaux}, {Fabricius}, {Khanna},
  {Muraveva}, {Reyl{\'e}}, {Spoto}, {Vallenari}, {Luri}, {Arenou}, {Alvarez},
  {Anders}, {Antoja}, {Balbinot}, {Barache}, {Bauchet}, {Bossini}, {Busonero},
  {Cantat-Gaudin}, {Carrasco}, {Dafonte}, {Diakite}, {Figueras},
  {Garcia-Gutierrez}, {Garofalo}, {Helmi}, {Jimenez-Arranz}, {Jordi},
  {Kervella}, {Kostrzewa-Rutkowska}, {Leclerc}, {Licata}, {Manteiga}, {Masip},
  {Monguio}, {Ramos}, {Robichon}, {Robin}, {Romero-Gomez}, {Saez}, {Santovena},
  {Spina}, {Torralba Elipe}, \& {Weiler}}]{Babusiaux2022arXiv220605989B}
{Babusiaux}, C., {Fabricius}, C., {Khanna}, S., {et~al.} 2022, arXiv e-prints,
  arXiv:2206.05989.
\newblock \doarXiv{2206.05989}

\bibitem[{{Bate}(2012)}]{Bate2012MNRAS.419.3115B}
{Bate}, M.~R. 2012, \mnras, 419, 3115, \dodoi{10.1111/j.1365-2966.2011.19955.x}

\bibitem[{{Bate} {et~al.}(2002){Bate}, {Bonnell}, \&
  {Bromm}}]{Bate2002MNRAS.336..705B}
{Bate}, M.~R., {Bonnell}, I.~A., \& {Bromm}, V. 2002, \mnras, 336, 705,
  \dodoi{10.1046/j.1365-8711.2002.05775.x}

\bibitem[{{Bate} {et~al.}(2003){Bate}, {Bonnell}, \&
  {Bromm}}]{Bate2003MNRAS.339..577B}
---. 2003, \mnras, 339, 577, \dodoi{10.1046/j.1365-8711.2003.06210.x}

\bibitem[{{B{\"o}hm-Vitense}(2006)}]{Bohm2006PASP..118..419B}
{B{\"o}hm-Vitense}, E. 2006, \pasp, 118, 419, \dodoi{10.1086/499385}

\bibitem[{{Bonnell} \& {Bate}(1994)}]{Bonnell1994}
{Bonnell}, I.~A., \& {Bate}, M.~R. 1994, \mnras, 269,
  \dodoi{10.1093/mnras/269.1.L45}

\bibitem[{{Borgniet} {et~al.}(2019){Borgniet}, {Lagrange}, {Meunier},
  {Galland}, {Arnold}, {Astudillo-Defru}, {Beuzit}, {Boisse}, {Bonfils},
  {Bouchy}, {Debondt}, {Deleuil}, {Delfosse}, {Desort}, {D{\'\i}az},
  {Eggenberger}, {Ehrenreich}, {Forveille}, {H{\'e}brard}, {Loeillet}, {Lovis},
  {Montagnier}, {Moutou}, {Pepe}, {Perrier}, {Pont}, {Queloz}, {Santerne},
  {Santos}, {S{\'e}gransan}, {da Silva}, {Sivan}, {Udry}, \&
  {Vidal-Madjar}}]{Borgniet2019}
{Borgniet}, S., {Lagrange}, A.~M., {Meunier}, N., {et~al.} 2019, \aap, 621,
  A87, \dodoi{10.1051/0004-6361/201833431}

\bibitem[{{Bourges} {et~al.}(2017){Bourges}, {Mella}, {Lafrasse}, {Duvert},
  {Chelli}, {Le Bouquin}, {Delfosse}, \&
  {Chesneau}}]{Bourges2017yCat.2346....0B}
{Bourges}, L., {Mella}, G., {Lafrasse}, S., {et~al.} 2017, VizieR Online Data
  Catalog, II/346

\bibitem[{{Burgasser} {et~al.}(2003){Burgasser}, {Kirkpatrick}, {Reid},
  {Brown}, {Miskey}, \& {Gizis}}]{Burg2003}
{Burgasser}, A.~J., {Kirkpatrick}, J.~D., {Reid}, I.~N., {et~al.} 2003, \apj,
  586, 512, \dodoi{10.1086/346263}

\bibitem[{{Carquillat} \& {Prieur}(2007)}]{Carquillat2007}
{Carquillat}, J.~M., \& {Prieur}, J.~L. 2007, \mnras, 380, 1064,
  \dodoi{10.1111/j.1365-2966.2007.12143.x}

\bibitem[{{Che} {et~al.}(2011){Che}, {Monnier}, {Zhao}, {Pedretti}, {Thureau},
  {M{\'e}rand}, {ten Brummelaar}, {McAlister}, {Ridgway}, {Turner}, {Sturmann},
  \& {Sturmann}}]{che2011ApJ...732...68C}
{Che}, X., {Monnier}, J.~D., {Zhao}, M., {et~al.} 2011, \apj, 732, 68,
  \dodoi{10.1088/0004-637X/732/2/68}

\bibitem[{{Choi} {et~al.}(2016){Choi}, {Dotter}, {Conroy}, {Cantiello},
  {Paxton}, \& {Johnson}}]{Choi2016ApJ...823..102C}
{Choi}, J., {Dotter}, A., {Conroy}, C., {et~al.} 2016, \apj, 823, 102,
  \dodoi{10.3847/0004-637X/823/2/102}

\bibitem[{{Cummings} {et~al.}(2018){Cummings}, {Kalirai}, {Tremblay},
  {Ramirez-Ruiz}, \& {Choi}}]{Cummings2018ApJ...866...21C}
{Cummings}, J.~D., {Kalirai}, J.~S., {Tremblay}, P.~E., {Ramirez-Ruiz}, E., \&
  {Choi}, J. 2018, \apj, 866, 21, \dodoi{10.3847/1538-4357/aadfd6}

\bibitem[{{De Furio} {et~al.}(2022){De Furio}, {Meyer}, {Reiter}, {Monnier},
  {Kraus}, \& {Dupuy}}]{DeFurio2022ApJ...925..112D}
{De Furio}, M., {Meyer}, M.~R., {Reiter}, M., {et~al.} 2022, \apj, 925, 112,
  \dodoi{10.3847/1538-4357/ac36d4}

\bibitem[{{De Furio} {et~al.}(2019){De Furio}, {Reiter}, {Meyer}, {Greenbaum},
  {Dupuy}, \& {Kraus}}]{DeFurio2019}
{De Furio}, M., {Reiter}, M., {Meyer}, M.~R., {et~al.} 2019, \apj, 886, 95,
  \dodoi{10.3847/1538-4357/ab4ae3}

\bibitem[{{De Rosa} {et~al.}(2014){De Rosa}, {Patience}, {Wilson}, {Schneider},
  {Wiktorowicz}, {Vigan}, {Marois}, {Song}, {Macintosh}, {Graham}, {Doyon},
  {Bessell}, {Thomas}, \& {Lai}}]{DeRosa2014}
{De Rosa}, R.~J., {Patience}, J., {Wilson}, P.~A., {et~al.} 2014, \mnras, 437,
  1216, \dodoi{10.1093/mnras/stt1932}

\bibitem[{{Dotter}(2016)}]{Dotter2016ApJS..222....8D}
{Dotter}, A. 2016, \apjs, 222, 8, \dodoi{10.3847/0067-0049/222/1/8}

\bibitem[{{Duch{\^e}ne} \& {Kraus}(2013)}]{DucheneKraus2013}
{Duch{\^e}ne}, G., \& {Kraus}, A. 2013, \araa, 51, 269,
  \dodoi{10.1146/annurev-astro-081710-102602}

\bibitem[{{El-Badry} {et~al.}(2019){El-Badry}, {Rix}, {Tian}, {Duch{\^e}ne}, \&
  {Moe}}]{ElBadry2019MNRAS.489.5822E}
{El-Badry}, K., {Rix}, H.-W., {Tian}, H., {Duch{\^e}ne}, G., \& {Moe}, M. 2019,
  \mnras, 489, 5822, \dodoi{10.1093/mnras/stz2480}

\bibitem[{{Gaia Collaboration} {et~al.}(2016){Gaia Collaboration}, {Prusti},
  {de Bruijne}, {Brown}, {Vallenari}, {Babusiaux}, {Bailer-Jones}, {Bastian},
  {Biermann}, {Evans}, {Eyer}, {Jansen}, {Jordi}, {Klioner}, {Lammers},
  {Lindegren}, {Luri}, {Mignard}, {Milligan}, {Panem}, {Poinsignon},
  {Pourbaix}, {Randich}, {Sarri}, {Sartoretti}, {Siddiqui}, {Soubiran},
  {Valette}, {van Leeuwen}, {Walton}, {Aerts}, {Arenou}, {Cropper}, {Drimmel},
  {H{\o}g}, {Katz}, {Lattanzi}, {O'Mullane}, {Grebel}, {Holland}, {Huc},
  {Passot}, {Bramante}, {Cacciari}, {Casta{\~n}eda}, {Chaoul}, {Cheek}, {De
  Angeli}, {Fabricius}, {Guerra}, {Hern{\'a}ndez}, {Jean-Antoine-Piccolo},
  {Masana}, {Messineo}, {Mowlavi}, {Nienartowicz}, {Ord{\'o}{\~n}ez-Blanco},
  {Panuzzo}, {Portell}, {Richards}, {Riello}, {Seabroke}, {Tanga},
  {Th{\'e}venin}, {Torra}, {Els}, {Gracia-Abril}, {Comoretto},
  {Garcia-Reinaldos}, {Lock}, {Mercier}, {Altmann}, {Andrae}, {Astraatmadja},
  {Bellas-Velidis}, {Benson}, {Berthier}, {Blomme}, {Busso}, {Carry},
  {Cellino}, {Clementini}, {Cowell}, {Creevey}, {Cuypers}, {Davidson}, {De
  Ridder}, {de Torres}, {Delchambre}, {Dell'Oro}, {Ducourant}, {Fr{\'e}mat},
  {Garc{\'\i}a-Torres}, {Gosset}, {Halbwachs}, {Hambly}, {Harrison}, {Hauser},
  {Hestroffer}, {Hodgkin}, {Huckle}, {Hutton}, {Jasniewicz}, {Jordan},
  {Kontizas}, {Korn}, {Lanzafame}, {Manteiga}, {Moitinho}, {Muinonen},
  {Osinde}, {Pancino}, {Pauwels}, {Petit}, {Recio-Blanco}, {Robin}, {Sarro},
  {Siopis}, {Smith}, {Smith}, {Sozzetti}, {Thuillot}, {van Reeven}, {Viala},
  {Abbas}, {Abreu Aramburu}, {Accart}, {Aguado}, {Allan}, {Allasia},
  {Altavilla}, {{\'A}lvarez}, {Alves}, {Anderson}, {Andrei}, {Anglada Varela},
  {Antiche}, {Antoja}, {Ant{\'o}n}, {Arcay}, {Atzei}, {Ayache}, {Bach},
  {Baker}, {Balaguer-N{\'u}{\~n}ez}, {Barache}, {Barata}, {Barbier}, {Barblan},
  {Baroni}, {Barrado y Navascu{\'e}s}, {Barros}, {Barstow}, {Becciani},
  {Bellazzini}, {Bellei}, {Bello Garc{\'\i}a}, {Belokurov}, {Bendjoya},
  {Berihuete}, {Bianchi}, {Bienaym{\'e}}, {Billebaud}, {Blagorodnova},
  {Blanco-Cuaresma}, {Boch}, {Bombrun}, {Borrachero}, {Bouquillon}, {Bourda},
  {Bouy}, {Bragaglia}, {Breddels}, {Brouillet}, {Br{\"u}semeister},
  {Bucciarelli}, {Budnik}, {Burgess}, {Burgon}, {Burlacu}, {Busonero}, {Buzzi},
  {Caffau}, {Cambras}, {Campbell}, {Cancelliere}, {Cantat-Gaudin}, {Carlucci},
  {Carrasco}, {Castellani}, {Charlot}, {Charnas}, {Charvet}, {Chassat},
  {Chiavassa}, {Clotet}, {Cocozza}, {Collins}, {Collins}, {Costigan}, {Crifo},
  {Cross}, {Crosta}, {Crowley}, {Dafonte}, {Damerdji}, {Dapergolas}, {David},
  {David}, {De Cat}, {de Felice}, {de Laverny}, {De Luise}, {De March}, {de
  Martino}, {de Souza}, {Debosscher}, {del Pozo}, {Delbo}, {Delgado},
  {Delgado}, {di Marco}, {Di Matteo}, {Diakite}, {Distefano}, {Dolding}, {Dos
  Anjos}, {Drazinos}, {Dur{\'a}n}, {Dzigan}, {Ecale}, {Edvardsson}, {Enke},
  {Erdmann}, {Escolar}, {Espina}, {Evans}, {Eynard Bontemps}, {Fabre},
  {Fabrizio}, {Faigler}, {Falc{\~a}o}, {Farr{\`a}s Casas}, {Faye}, {Federici},
  {Fedorets}, {Fern{\'a}ndez-Hern{\'a}ndez}, {Fernique}, {Fienga}, {Figueras},
  {Filippi}, {Findeisen}, {Fonti}, {Fouesneau}, {Fraile}, {Fraser}, {Fuchs},
  {Furnell}, {Gai}, {Galleti}, {Galluccio}, {Garabato}, {Garc{\'\i}a-Sedano},
  {Gar{\'e}}, {Garofalo}, {Garralda}, {Gavras}, {Gerssen}, {Geyer}, {Gilmore},
  {Girona}, {Giuffrida}, {Gomes}, {Gonz{\'a}lez-Marcos},
  {Gonz{\'a}lez-N{\'u}{\~n}ez}, {Gonz{\'a}lez-Vidal}, {Granvik}, {Guerrier},
  {Guillout}, {Guiraud}, {G{\'u}rpide}, {Guti{\'e}rrez-S{\'a}nchez}, {Guy},
  {Haigron}, {Hatzidimitriou}, {Haywood}, {Heiter}, {Helmi}, {Hobbs},
  {Hofmann}, {Holl}, {Holland}, {Hunt}, {Hypki}, {Icardi}, {Irwin}, {Jevardat
  de Fombelle}, {Jofr{\'e}}, {Jonker}, {Jorissen}, {Julbe}, {Karampelas},
  {Kochoska}, {Kohley}, {Kolenberg}, {Kontizas}, {Koposov}, {Kordopatis},
  {Koubsky}, {Kowalczyk}, {Krone-Martins}, {Kudryashova}, {Kull}, {Bachchan},
  {Lacoste-Seris}, {Lanza}, {Lavigne}, {Le Poncin-Lafitte}, {Lebreton},
  {Lebzelter}, {Leccia}, {Leclerc}, {Lecoeur-Taibi}, {Lemaitre}, {Lenhardt},
  {Leroux}, {Liao}, {Licata}, {Lindstr{\o}m}, {Lister}, {Livanou}, {Lobel},
  {L{\"o}ffler}, {L{\'o}pez}, {Lopez-Lozano}, {Lorenz}, {Loureiro},
  {MacDonald}, {Magalh{\~a}es Fernandes}, {Managau}, {Mann}, {Mantelet},
  {Marchal}, {Marchant}, {Marconi}, {Marie}, {Marinoni}, {Marrese},
  {Marschalk{\'o}}, {Marshall}, {Mart{\'\i}n-Fleitas}, {Martino}, {Mary},
  {Matijevi{\v{c}}}, {Mazeh}, {McMillan}, {Messina}, {Mestre}, {Michalik},
  {Millar}, {Miranda}, {Molina}, {Molinaro}, {Molinaro}, {Moln{\'a}r},
  {Moniez}, {Montegriffo}, {Monteiro}, {Mor}, {Mora}, {Morbidelli}, {Morel},
  {Morgenthaler}, {Morley}, {Morris}, {Mulone}, {Muraveva}, {Musella},
  {Narbonne}, {Nelemans}, {Nicastro}, {Noval}, {Ord{\'e}novic},
  {Ordieres-Mer{\'e}}, {Osborne}, {Pagani}, {Pagano}, {Pailler}, {Palacin},
  {Palaversa}, {Parsons}, {Paulsen}, {Pecoraro}, {Pedrosa}, {Pentik{\"a}inen},
  {Pereira}, {Pichon}, {Piersimoni}, {Pineau}, {Plachy}, {Plum}, {Poujoulet},
  {Pr{\v{s}}a}, {Pulone}, {Ragaini}, {Rago}, {Rambaux}, {Ramos-Lerate},
  {Ranalli}, {Rauw}, {Read}, {Regibo}, {Renk}, {Reyl{\'e}}, {Ribeiro},
  {Rimoldini}, {Ripepi}, {Riva}, {Rixon}, {Roelens}, {Romero-G{\'o}mez},
  {Rowell}, {Royer}, {Rudolph}, {Ruiz-Dern}, {Sadowski}, {Sagrist{\`a}
  Sell{\'e}s}, {Sahlmann}, {Salgado}, {Salguero}, {Sarasso}, {Savietto},
  {Schnorhk}, {Schultheis}, {Sciacca}, {Segol}, {Segovia}, {Segransan},
  {Serpell}, {Shih}, {Smareglia}, {Smart}, {Smith}, {Solano}, {Solitro},
  {Sordo}, {Soria Nieto}, {Souchay}, {Spagna}, {Spoto}, {Stampa}, {Steele},
  {Steidelm{\"u}ller}, {Stephenson}, {Stoev}, {Suess}, {S{\"u}veges}, {Surdej},
  {Szabados}, {Szegedi-Elek}, {Tapiador}, {Taris}, {Tauran}, {Taylor},
  {Teixeira}, {Terrett}, {Tingley}, {Trager}, {Turon}, {Ulla}, {Utrilla},
  {Valentini}, {van Elteren}, {Van Hemelryck}, {van Leeuwen}, {Varadi},
  {Vecchiato}, {Veljanoski}, {Via}, {Vicente}, {Vogt}, {Voss}, {Votruba},
  {Voutsinas}, {Walmsley}, {Weiler}, {Weingrill}, {Werner}, {Wevers},
  {Whitehead}, {Wyrzykowski}, {Yoldas}, {{\v{Z}}erjal}, {Zucker}, {Zurbach},
  {Zwitter}, {Alecu}, {Allen}, {Allende Prieto}, {Amorim},
  {Anglada-Escud{\'e}}, {Arsenijevic}, {Azaz}, {Balm}, {Beck}, {Bernstein},
  {Bigot}, {Bijaoui}, {Blasco}, {Bonfigli}, {Bono}, {Boudreault}, {Bressan},
  {Brown}, {Brunet}, {Bunclark}, {Buonanno}, {Butkevich}, {Carret}, {Carrion},
  {Chemin}, {Ch{\'e}reau}, {Corcione}, {Darmigny}, {de Boer}, {de Teodoro}, {de
  Zeeuw}, {Delle Luche}, {Domingues}, {Dubath}, {Fodor}, {Fr{\'e}zouls},
  {Fries}, {Fustes}, {Fyfe}, {Gallardo}, {Gallegos}, {Gardiol}, {Gebran},
  {Gomboc}, {G{\'o}mez}, {Grux}, {Gueguen}, {Heyrovsky}, {Hoar}, {Iannicola},
  {Isasi Parache}, {Janotto}, {Joliet}, {Jonckheere}, {Keil}, {Kim},
  {Klagyivik}, {Klar}, {Knude}, {Kochukhov}, {Kolka}, {Kos}, {Kutka}, {Lainey},
  {LeBouquin}, {Liu}, {Loreggia}, {Makarov}, {Marseille}, {Martayan},
  {Martinez-Rubi}, {Massart}, {Meynadier}, {Mignot}, {Munari}, {Nguyen},
  {Nordlander}, {Ocvirk}, {O'Flaherty}, {Olias Sanz}, {Ortiz}, {Osorio},
  {Oszkiewicz}, {Ouzounis}, {Palmer}, {Park}, {Pasquato}, {Peltzer}, {Peralta},
  {P{\'e}turaud}, {Pieniluoma}, {Pigozzi}, {Poels}, {Prat}, {Prod'homme},
  {Raison}, {Rebordao}, {Risquez}, {Rocca-Volmerange}, {Rosen}, {Ruiz-Fuertes},
  {Russo}, {Sembay}, {Serraller Vizcaino}, {Short}, {Siebert}, {Silva},
  {Sinachopoulos}, {Slezak}, {Soffel}, {Sosnowska}, {Strai{\v{z}}ys}, {ter
  Linden}, {Terrell}, {Theil}, {Tiede}, {Troisi}, {Tsalmantza}, {Tur},
  {Vaccari}, {Vachier}, {Valles}, {Van Hamme}, {Veltz}, {Virtanen}, {Wallut},
  {Wichmann}, {Wilkinson}, {Ziaeepour}, \& {Zschocke}}]{TheGaiaMission2016}
{Gaia Collaboration}, {Prusti}, T., {de Bruijne}, J.~H.~J., {et~al.} 2016,
  \aap, 595, A1, \dodoi{10.1051/0004-6361/201629272}

\bibitem[{{Gaia Collaboration} {et~al.}(2021){Gaia Collaboration}, {Brown},
  {Vallenari}, {Prusti}, {de Bruijne}, {Babusiaux}, {Biermann}, {Creevey},
  {Evans}, {Eyer}, {Hutton}, {Jansen}, {Jordi}, {Klioner}, {Lammers},
  {Lindegren}, {Luri}, {Mignard}, {Panem}, {Pourbaix}, {Randich}, {Sartoretti},
  {Soubiran}, {Walton}, {Arenou}, {Bailer-Jones}, {Bastian}, {Cropper},
  {Drimmel}, {Katz}, {Lattanzi}, {van Leeuwen}, {Bakker}, {Cacciari},
  {Casta{\~n}eda}, {De Angeli}, {Ducourant}, {Fabricius}, {Fouesneau},
  {Fr{\'e}mat}, {Guerra}, {Guerrier}, {Guiraud}, {Jean-Antoine Piccolo},
  {Masana}, {Messineo}, {Mowlavi}, {Nicolas}, {Nienartowicz}, {Pailler},
  {Panuzzo}, {Riclet}, {Roux}, {Seabroke}, {Sordo}, {Tanga}, {Th{\'e}venin},
  {Gracia-Abril}, {Portell}, {Teyssier}, {Altmann}, {Andrae}, {Bellas-Velidis},
  {Benson}, {Berthier}, {Blomme}, {Brugaletta}, {Burgess}, {Busso}, {Carry},
  {Cellino}, {Cheek}, {Clementini}, {Damerdji}, {Davidson}, {Delchambre},
  {Dell'Oro}, {Fern{\'a}ndez-Hern{\'a}ndez}, {Galluccio}, {Garc{\'\i}a-Lario},
  {Garcia-Reinaldos}, {Gonz{\'a}lez-N{\'u}{\~n}ez}, {Gosset}, {Haigron},
  {Halbwachs}, {Hambly}, {Harrison}, {Hatzidimitriou}, {Heiter},
  {Hern{\'a}ndez}, {Hestroffer}, {Hodgkin}, {Holl}, {Jan{\ss}en}, {Jevardat de
  Fombelle}, {Jordan}, {Krone-Martins}, {Lanzafame}, {L{\"o}ffler}, {Lorca},
  {Manteiga}, {Marchal}, {Marrese}, {Moitinho}, {Mora}, {Muinonen}, {Osborne},
  {Pancino}, {Pauwels}, {Petit}, {Recio-Blanco}, {Richards}, {Riello},
  {Rimoldini}, {Robin}, {Roegiers}, {Rybizki}, {Sarro}, {Siopis}, {Smith},
  {Sozzetti}, {Ulla}, {Utrilla}, {van Leeuwen}, {van Reeven}, {Abbas}, {Abreu
  Aramburu}, {Accart}, {Aerts}, {Aguado}, {Ajaj}, {Altavilla}, {{\'A}lvarez},
  {{\'A}lvarez Cid-Fuentes}, {Alves}, {Anderson}, {Anglada Varela}, {Antoja},
  {Audard}, {Baines}, {Baker}, {Balaguer-N{\'u}{\~n}ez}, {Balbinot}, {Balog},
  {Barache}, {Barbato}, {Barros}, {Barstow}, {Bartolom{\'e}}, {Bassilana},
  {Bauchet}, {Baudesson-Stella}, {Becciani}, {Bellazzini}, {Bernet}, {Bertone},
  {Bianchi}, {Blanco-Cuaresma}, {Boch}, {Bombrun}, {Bossini}, {Bouquillon},
  {Bragaglia}, {Bramante}, {Breedt}, {Bressan}, {Brouillet}, {Bucciarelli},
  {Burlacu}, {Busonero}, {Butkevich}, {Buzzi}, {Caffau}, {Cancelliere},
  {C{\'a}novas}, {Cantat-Gaudin}, {Carballo}, {Carlucci}, {Carnerero},
  {Carrasco}, {Casamiquela}, {Castellani}, {Castro-Ginard}, {Castro Sampol},
  {Chaoul}, {Charlot}, {Chemin}, {Chiavassa}, {Cioni}, {Comoretto}, {Cooper},
  {Cornez}, {Cowell}, {Crifo}, {Crosta}, {Crowley}, {Dafonte}, {Dapergolas},
  {David}, {David}, {de Laverny}, {De Luise}, {De March}, {De Ridder}, {de
  Souza}, {de Teodoro}, {de Torres}, {del Peloso}, {del Pozo}, {Delbo},
  {Delgado}, {Delgado}, {Delisle}, {Di Matteo}, {Diakite}, {Diener},
  {Distefano}, {Dolding}, {Eappachen}, {Edvardsson}, {Enke}, {Esquej}, {Fabre},
  {Fabrizio}, {Faigler}, {Fedorets}, {Fernique}, {Fienga}, {Figueras},
  {Fouron}, {Fragkoudi}, {Fraile}, {Franke}, {Gai}, {Garabato},
  {Garcia-Gutierrez}, {Garc{\'\i}a-Torres}, {Garofalo}, {Gavras}, {Gerlach},
  {Geyer}, {Giacobbe}, {Gilmore}, {Girona}, {Giuffrida}, {Gomel}, {Gomez},
  {Gonzalez-Santamaria}, {Gonz{\'a}lez-Vidal}, {Granvik},
  {Guti{\'e}rrez-S{\'a}nchez}, {Guy}, {Hauser}, {Haywood}, {Helmi}, {Hidalgo},
  {Hilger}, {H{\l}adczuk}, {Hobbs}, {Holland}, {Huckle}, {Jasniewicz},
  {Jonker}, {Juaristi Campillo}, {Julbe}, {Karbevska}, {Kervella}, {Khanna},
  {Kochoska}, {Kontizas}, {Kordopatis}, {Korn}, {Kostrzewa-Rutkowska},
  {Kruszy{\'n}ska}, {Lambert}, {Lanza}, {Lasne}, {Le Campion}, {Le Fustec},
  {Lebreton}, {Lebzelter}, {Leccia}, {Leclerc}, {Lecoeur-Taibi}, {Liao},
  {Licata}, {Lindstr{\o}m}, {Lister}, {Livanou}, {Lobel}, {Madrero Pardo},
  {Managau}, {Mann}, {Marchant}, {Marconi}, {Marcos Santos}, {Marinoni},
  {Marocco}, {Marshall}, {Martin Polo}, {Mart{\'\i}n-Fleitas}, {Masip},
  {Massari}, {Mastrobuono-Battisti}, {Mazeh}, {McMillan}, {Messina},
  {Michalik}, {Millar}, {Mints}, {Molina}, {Molinaro}, {Moln{\'a}r},
  {Montegriffo}, {Mor}, {Morbidelli}, {Morel}, {Morris}, {Mulone}, {Munoz},
  {Muraveva}, {Murphy}, {Musella}, {Noval}, {Ord{\'e}novic}, {Orr{\`u}},
  {Osinde}, {Pagani}, {Pagano}, {Palaversa}, {Palicio}, {Panahi}, {Pawlak},
  {Pe{\~n}alosa Esteller}, {Penttil{\"a}}, {Piersimoni}, {Pineau}, {Plachy},
  {Plum}, {Poggio}, {Poretti}, {Poujoulet}, {Pr{\v{s}}a}, {Pulone}, {Racero},
  {Ragaini}, {Rainer}, {Raiteri}, {Rambaux}, {Ramos}, {Ramos-Lerate}, {Re
  Fiorentin}, {Regibo}, {Reyl{\'e}}, {Ripepi}, {Riva}, {Rixon}, {Robichon},
  {Robin}, {Roelens}, {Rohrbasser}, {Romero-G{\'o}mez}, {Rowell}, {Royer},
  {Rybicki}, {Sadowski}, {Sagrist{\`a} Sell{\'e}s}, {Sahlmann}, {Salgado},
  {Salguero}, {Samaras}, {Sanchez Gimenez}, {Sanna}, {Santove{\~n}a},
  {Sarasso}, {Schultheis}, {Sciacca}, {Segol}, {Segovia}, {S{\'e}gransan},
  {Semeux}, {Shahaf}, {Siddiqui}, {Siebert}, {Siltala}, {Slezak}, {Smart},
  {Solano}, {Solitro}, {Souami}, {Souchay}, {Spagna}, {Spoto}, {Steele},
  {Steidelm{\"u}ller}, {Stephenson}, {S{\"u}veges}, {Szabados}, {Szegedi-Elek},
  {Taris}, {Tauran}, {Taylor}, {Teixeira}, {Thuillot}, {Tonello}, {Torra},
  {Torra}, {Turon}, {Unger}, {Vaillant}, {van Dillen}, {Vanel}, {Vecchiato},
  {Viala}, {Vicente}, {Voutsinas}, {Weiler}, {Wevers}, {Wyrzykowski}, {Yoldas},
  {Yvard}, {Zhao}, {Zorec}, {Zucker}, {Zurbach}, \&
  {Zwitter}}]{earlygaia2021A&A...649A...1G}
{Gaia Collaboration}, {Brown}, A.~G.~A., {Vallenari}, A., {et~al.} 2021, \aap,
  649, A1, \dodoi{10.1051/0004-6361/202039657}

\bibitem[{{Gaia Collaboration} {et~al.}(2022){Gaia Collaboration}, {Arenou},
  {Babusiaux}, {Barstow}, {Faigler}, {Jorissen}, {Kervella}, {Mazeh},
  {Mowlavi}, {Panuzzo}, {Sahlmann}, {Shahaf}, {Sozzetti}, {Bauchet},
  {Damerdji}, {Gavras}, {Giacobbe}, {Gosset}, {Halbwachs}, {Holl}, {Lattanzi},
  {Leclerc}, {Morel}, {Pourbaix}, {Re Fiorentin}, {Sadowski}, {S{\'e}gransan},
  {Siopis}, {Teyssier}, {Zwitter}, {Planquart}, {Brown}, {Vallenari}, {Prusti},
  {de Bruijne}, {Biermann}, {Creevey}, {Ducourant}, {Evans}, {Eyer}, {Guerra},
  {Hutton}, {Jordi}, {Klioner}, {Lammers}, {Lindegren}, {Luri}, {Mignard},
  {Panem}, {Randich}, {Sartoretti}, {Soubiran}, {Tanga}, {Walton},
  {Bailer-Jones}, {Bastian}, {Drimmel}, {Jansen}, {Katz}, {van Leeuwen},
  {Bakker}, {Cacciari}, {Casta{\~n}eda}, {De Angeli}, {Fabricius}, {Fouesneau},
  {Fr{\'e}mat}, {Galluccio}, {Guerrier}, {Heiter}, {Masana}, {Messineo},
  {Nicolas}, {Nienartowicz}, {Pailler}, {Riclet}, {Roux}, {Seabroke}, {Sordo},
  {Th{\'e}venin}, {Gracia-Abril}, {Portell}, {Altmann}, {Andrae}, {Audard},
  {Bellas-Velidis}, {Benson}, {Berthier}, {Blomme}, {Burgess}, {Busonero},
  {Busso}, {C{\'a}novas}, {Carry}, {Cellino}, {Cheek}, {Clementini},
  {Davidson}, {de Teodoro}, {Nu{\~n}ez Campos}, {Delchambre}, {Dell'Oro},
  {Esquej}, {Fern{\'a}ndez-Hern{\'a}ndez}, {Fraile}, {Garabato},
  {Garc{\'\i}a-Lario}, {Haigron}, {Hambly}, {Harrison}, {Hern{\'a}ndez},
  {Hestroffer}, {Hodgkin}, {Jan{\ss}en}, {Jevardat de Fombelle}, {Jordan},
  {Krone-Martins}, {Lanzafame}, {L{\"o}ffler}, {Marchal}, {Marrese},
  {Moitinho}, {Muinonen}, {Osborne}, {Pancino}, {Pauwels}, {Recio-Blanco},
  {Reyl{\'e}}, {Riello}, {Rimoldini}, {Roegiers}, {Rybizki}, {Sarro}, {Smith},
  {Utrilla}, {van Leeuwen}, {Abbas}, {{\'A}brah{\'a}m}, {Abreu Aramburu},
  {Aerts}, {Aguado}, {Ajaj}, {Aldea-Montero}, {Altavilla}, {{\'A}lvarez},
  {Alves}, {Anders}, {Anderson}, {Anglada Varela}, {Antoja}, {Baines}, {Baker},
  {Balaguer-N{\'u}{\~n}ez}, {Balbinot}, {Balog}, {Barache}, {Barbato},
  {Barros}, {Bartolom{\'e}}, {Bassilana}, {Becciani}, {Bellazzini},
  {Berihuete}, {Bernet}, {Bertone}, {Bianchi}, {Binnenfeld}, {Blanco-Cuaresma},
  {Blazere}, {Boch}, {Bombrun}, {Bossini}, {Bouquillon}, {Bragaglia},
  {Bramante}, {Breedt}, {Bressan}, {Brouillet}, {Brugaletta}, {Bucciarelli},
  {Burlacu}, {Butkevich}, {Buzzi}, {Caffau}, {Cancelliere}, {Cantat-Gaudin},
  {Carballo}, {Carlucci}, {Carnerero}, {Carrasco}, {Casamiquela}, {Castellani},
  {Castro-Ginard}, {Chaoul}, {Charlot}, {Chemin}, {Chiaramida}, {Chiavassa},
  {Chornay}, {Comoretto}, {Contursi}, {Cooper}, {Cornez}, {Cowell}, {Crifo},
  {Cropper}, {Crosta}, {Crowley}, {Dafonte}, {Dapergolas}, {David}, {de
  Laverny}, {De Luise}, {De March}, {De Ridder}, {de Souza}, {de Torres}, {del
  Peloso}, {del Pozo}, {Delbo}, {Delgado}, {Delisle}, {Demouchy},
  {Dharmawardena}, {Diakite}, {Diener}, {Distefano}, {Dolding}, {Enke},
  {Fabre}, {Fabrizio}, {Fedorets}, {Fernique}, {Figueras}, {Fournier},
  {Fouron}, {Fragkoudi}, {Gai}, {Garcia-Gutierrez}, {Garcia-Reinaldos},
  {Garc{\'\i}a-Torres}, {Garofalo}, {Gavel}, {Gerlach}, {Geyer}, {Gilmore},
  {Girona}, {Giuffrida}, {Gomel}, {Gomez}, {Gonz{\'a}lez-N{\'u}{\~n}ez},
  {Gonz{\'a}lez-Santamar{\'\i}a}, {Gonz{\'a}lez-Vidal}, {Granvik}, {Guillout},
  {Guiraud}, {Guti{\'e}rrez-S{\'a}nchez}, {Guy}, {Hatzidimitriou}, {Hauser},
  {Haywood}, {Helmer}, {Helmi}, {Sarmiento}, {Hidalgo}, {H{\l}adczuk}, {Hobbs},
  {Holland}, {Huckle}, {Jardine}, {Jasniewicz}, {Jean-Antoine Piccolo},
  {Jim{\'e}nez-Arranz}, {Juaristi Campillo}, {Julbe}, {Karbevska}, {Khanna},
  {Kordopatis}, {Korn}, {K{\'o}sp{\'a}l}, {Kostrzewa-Rutkowska},
  {Kruszy{\'n}ska}, {Kun}, {Laizeau}, {Lambert}, {Lanza}, {Lasne}, {Le
  Campion}, {Lebreton}, {Lebzelter}, {Leccia}, {Lecoeur-Taibi}, {Liao},
  {Licata}, {Lindstr{\o}m}, {Lister}, {Livanou}, {Lobel}, {Lorca}, {Loup},
  {Madrero Pardo}, {Magdaleno Romeo}, {Managau}, {Mann}, {Manteiga},
  {Marchant}, {Marconi}, {Marcos}, {Marcos Santos}, {Mar{\'\i}n Pina},
  {Marinoni}, {Marocco}, {Marshall}, {Polo}, {Mart{\'\i}n-Fleitas}, {Marton},
  {Mary}, {Masip}, {Massari}, {Mastrobuono-Battisti}, {McMillan}, {Messina},
  {Michalik}, {Millar}, {Mints}, {Molina}, {Molinaro}, {Moln{\'a}r}, {Monari},
  {Mongui{\'o}}, {Montegriffo}, {Montero}, {Mor}, {Mora}, {Morbidelli},
  {Morris}, {Muraveva}, {Murphy}, {Musella}, {Nagy}, {Noval}, {Oca{\~n}a},
  {Ogden}, {Ordenovic}, {Osinde}, {Pagani}, {Pagano}, {Palaversa}, {Palicio},
  {Pallas-Quintela}, {Panahi}, {Payne-Wardenaar}, {Pe{\~n}alosa Esteller},
  {Penttil{\"a}}, {Pichon}, {Piersimoni}, {Pineau}, {Plachy}, {Plum}, {Poggio},
  {Pr{\v{s}}a}, {Pulone}, {Racero}, {Ragaini}, {Rainer}, {Raiteri}, {Ramos},
  {Ramos-Lerate}, {Regibo}, {Richards}, {Rios Diaz}, {Ripepi}, {Riva}, {Rix},
  {Rixon}, {Robichon}, {Robin}, {Robin}, {Roelens}, {Rogues}, {Rohrbasser},
  {Romero-G{\'o}mez}, {Rowell}, {Royer}, {Ruz Mieres}, {Rybicki}, {S{\'a}ez
  N{\'u}{\~n}ez}, {Sagrist{\`a} Sell{\'e}s}, {Salguero}, {Samaras}, {Sanchez
  Gimenez}, {Sanna}, {Santove{\~n}a}, {Sarasso}, {Schultheis}, {Sciacca},
  {Segol}, {Segovia}, {Semeux}, {Siddiqui}, {Siebert}, {Siltala}, {Silvelo},
  {Slezak}, {Slezak}, {Smart}, {Snaith}, {Solano}, {Solitro}, {Souami},
  {Souchay}, {Spagna}, {Spina}, {Spoto}, {Steele}, {Steidelm{\"u}ller},
  {Stephenson}, {S{\"u}veges}, {Surdej}, {Szabados}, {Szegedi-Elek}, {Taris},
  {Taylor}, {Teixeira}, {Tolomei}, {Tonello}, {Torra}, {Torra}, {Torralba
  Elipe}, {Trabucchi}, {Tsounis}, {Turon}, {Ulla}, {Unger}, {Vaillant}, {van
  Dillen}, {van Reeven}, {Vanel}, {Vecchiato}, {Viala}, {Vicente}, {Voutsinas},
  {Weiler}, {Wevers}, {Wyrzykowski}, {Yoldas}, {Yvard}, {Zhao}, {Zorec}, \&
  {Zucker}}]{Gaia_Multiplicity_2022arXiv220605595G}
{Gaia Collaboration}, {Arenou}, F., {Babusiaux}, C., {et~al.} 2022, arXiv
  e-prints, arXiv:2206.05595.
\newblock \doarXiv{2206.05595}

\bibitem[{{Gallenne} {et~al.}(2015){Gallenne}, {M{\'e}rand}, {Kervella},
  {Monnier}, {Schaefer}, {Baron}, {Breitfelder}, {Le Bouquin}, {Roettenbacher},
  {Gieren}, {Pietrzy{\'n}ski}, {McAlister}, {ten Brummelaar}, {Sturmann},
  {Sturmann}, {Turner}, {Ridgway}, \& {Kraus}}]{Gallenne2015}
{Gallenne}, A., {M{\'e}rand}, A., {Kervella}, P., {et~al.} 2015, \aap, 579,
  A68, \dodoi{10.1051/0004-6361/201525917}

\bibitem[{{Goodwin} {et~al.}(2004){Goodwin}, {Whitworth}, \&
  {Ward-Thompson}}]{Goodwin2004}
{Goodwin}, S.~P., {Whitworth}, A.~P., \& {Ward-Thompson}, D. 2004, \aap, 414,
  633, \dodoi{10.1051/0004-6361:20031594}

\bibitem[{{Gravity Collaboration} {et~al.}(2018){Gravity Collaboration},
  {Karl}, {Pfuhl}, {Eisenhauer}, {Genzel}, {Grellmann}, {Habibi}, {Abuter},
  {Accardo}, {Amorim}, {Anugu}, {{\'A}vila}, {Benisty}, {Berger}, {Blind},
  {Bonnet}, {Bourget}, {Brandner}, {Brast}, {Buron}, {Caratti O Garatti},
  {Chapron}, {Cl{\'e}net}, {Collin}, {Coud{\'e} Du Foresto}, {de Wit}, {de
  Zeeuw}, {Deen}, {Delplancke-Str{\"o}bele}, {Dembet}, {Derie}, {Dexter},
  {Duvert}, {Ebert}, {Eckart}, {Esselborn}, {F{\'e}dou}, {Finger}, {Garcia},
  {Garcia Dabo}, {Garcia Lopez}, {Gao}, {Gendron}, {Gillessen}, {Gont{\'e}},
  {Gordo}, {Gr{\"o}zinger}, {Guajardo}, {Guieu}, {Haguenauer}, {Hans},
  {Haubois}, {Haug}, {Hau{\ss}mann}, {Henning}, {Hippler}, {Horrobin}, {Huber},
  {Hubert}, {Hubin}, {Jakob}, {Jochum}, {Jocou}, {Kaufer}, {Kellner},
  {Kendrew}, {Kern}, {Kervella}, {Kiekebusch}, {Klein}, {K{\"o}hler}, {Kolb},
  {Kulas}, {Lacour}, {Lapeyr{\`e}re}, {Lazareff}, {Le Bouquin}, {L{\'e}na},
  {Lenzen}, {L{\'e}v{\^e}que}, {Lin}, {Lippa}, {Magnard}, {Mehrgan},
  {M{\'e}rand}, {Moulin}, {M{\"u}ller}, {M{\"u}ller}, {Neumann}, {Oberti},
  {Ott}, {Pallanca}, {Panduro}, {Pasquini}, {Paumard}, {Percheron}, {Perraut},
  {Perrin}, {Pfl{\"u}ger}, {Duc}, {Plewa}, {Popovic}, {Rabien}, {Ram{\'\i}rez},
  {Ramos}, {Rau}, {Riquelme}, {Rodr{\'\i}guez-Coira}, {Rohloff}, {Rosales},
  {Rousset}, {Sanchez-Bermudez}, {Scheithauer}, {Sch{\"o}ller}, {Schuhler},
  {Spyromilio}, {Straub}, {Straubmeier}, {Sturm}, {Suarez}, {Tristram},
  {Ventura}, {Vincent}, {Waisberg}, {Wank}, {Widmann}, {Wieprecht}, {Wiest},
  {Wiezorrek}, {Wittkowski}, {Woillez}, {Wolff}, {Yazici}, {Ziegler}, \&
  {Zins}}]{Karl2018}
{Gravity Collaboration}, {Karl}, M., {Pfuhl}, O., {et~al.} 2018, \aap, 620,
  A116, \dodoi{10.1051/0004-6361/201833575}

\bibitem[{{Gray} {et~al.}(2016){Gray}, {Corbally}, {De Cat}, {Fu}, {Ren},
  {Shi}, {Luo}, {Zhang}, {Wu}, {Cao}, {Li}, {Zhang}, {Hou}, \&
  {Wang}}]{Gray2016AJ....151...13G}
{Gray}, R.~O., {Corbally}, C.~J., {De Cat}, P., {et~al.} 2016, \aj, 151, 13,
  \dodoi{10.3847/0004-6256/151/1/13}

\bibitem[{{H{\"u}mmerich} {et~al.}(2017){H{\"u}mmerich}, {Bernhard}, {Paunzen},
  {Hambsch}, {Bohlsen}, \& {Powles}}]{Hummerich2017MNRAS.466.1399H}
{H{\"u}mmerich}, S., {Bernhard}, K., {Paunzen}, E., {et~al.} 2017, \mnras, 466,
  1399, \dodoi{10.1093/mnras/stw3186}

\bibitem[{{Janson} {et~al.}(2012){Janson}, {Hormuth}, {Bergfors}, {Brandner},
  {Hippler}, {Daemgen}, {Kudryavtseva}, {Schmalzl}, {Schnupp}, \&
  {Henning}}]{Janson2012}
{Janson}, M., {Hormuth}, F., {Bergfors}, C., {et~al.} 2012, \apj, 754, 44,
  \dodoi{10.1088/0004-637X/754/1/44}

\bibitem[{{Jennison}(1958)}]{Jennison1958MNRAS.118..276J}
{Jennison}, R.~C. 1958, \mnras, 118, 276, \dodoi{10.1093/mnras/118.3.276}

\bibitem[{{Kervella} {et~al.}(2019){Kervella}, {Arenou}, {Mignard}, \&
  {Th{\'e}venin}}]{Kervella2019A&A...623A..72K}
{Kervella}, P., {Arenou}, F., {Mignard}, F., \& {Th{\'e}venin}, F. 2019, \aap,
  623, A72, \dodoi{10.1051/0004-6361/201834371}

\bibitem[{{Kratter}(2011)}]{Kratter2011ASPC..447...47K}
{Kratter}, K.~M. 2011, in Astronomical Society of the Pacific Conference
  Series, Vol. 447, Evolution of Compact Binaries, ed. L.~{Schmidtobreick},
  M.~R. {Schreiber}, \& C.~{Tappert}, 47.
\newblock \doarXiv{1109.3740}

\bibitem[{{Kratter} {et~al.}(2008){Kratter}, {Matzner}, \&
  {Krumholz}}]{Kratter2008}
{Kratter}, K.~M., {Matzner}, C.~D., \& {Krumholz}, M.~R. 2008, \apj, 681, 375,
  \dodoi{10.1086/587543}

\bibitem[{{Lee} {et~al.}(2019){Lee}, {Offner}, {Kratter}, {Smullen}, \&
  {Li}}]{Lee2019ApJ...887..232L}
{Lee}, A.~T., {Offner}, S. S.~R., {Kratter}, K.~M., {Smullen}, R.~A., \& {Li},
  P.~S. 2019, \apj, 887, 232, \dodoi{10.3847/1538-4357/ab584b}

\bibitem[{{Michaud}(1980)}]{Michaud1980AJ.....85..589M}
{Michaud}, G. 1980, \aj, 85, 589, \dodoi{10.1086/112716}

\bibitem[{{Michaud} {et~al.}(1983){Michaud}, {Tarasick}, {Charland}, \&
  {Pelletier}}]{Michaud1983ApJ...269..239M}
{Michaud}, G., {Tarasick}, D., {Charland}, Y., \& {Pelletier}, C. 1983, \apj,
  269, 239, \dodoi{10.1086/161034}

\bibitem[{{Moe} \& {Di Stefano}(2017)}]{Moe2017}
{Moe}, M., \& {Di Stefano}, R. 2017, \apjs, 230, 15,
  \dodoi{10.3847/1538-4365/aa6fb6}

\bibitem[{{Moe} \& {Kratter}(2018)}]{MoeKratter2018}
{Moe}, M., \& {Kratter}, K.~M. 2018, \apj, 854, 44,
  \dodoi{10.3847/1538-4357/aaa6d2}

\bibitem[{{Monnier}(2000)}]{Monnier2000plbs.conf..203M}
{Monnier}, J.~D. 2000, in Principles of Long Baseline Stellar Interferometry,
  ed. P.~R. {Lawson}, 203

\bibitem[{{Monnier} {et~al.}(2004){Monnier}, {Traub}, {Schloerb},
  {Millan-Gabet}, {Berger}, {Pedretti}, {Carleton}, {Kraus}, {Lacasse},
  {Brewer}, {Ragland}, {Ahearn}, {Coldwell}, {Haguenauer}, {Kern}, {Labeye},
  {Lagny}, {Malbet}, {Malin}, {Maymounkov}, {Morel}, {Papaliolios}, {Perraut},
  {Pearlman}, {Porro}, {Schanen}, {Souccar}, {Torres}, \&
  {Wallace}}]{Monnier2004ApJ...602L..57M}
{Monnier}, J.~D., {Traub}, W.~A., {Schloerb}, F.~P., {et~al.} 2004, \apjl, 602,
  L57, \dodoi{10.1086/382213}

\bibitem[{{Monnier} {et~al.}(2006){Monnier}, {Pedretti}, {Thureau}, {Berger},
  {Millan-Gabet}, {ten Brummelaar}, {McAlister}, {Sturmann}, {Sturmann},
  {Muirhead}, {Tannirkulam}, {Webster}, \& {Zhao}}]{Monnier2006SPIE.6268E..1PM}
{Monnier}, J.~D., {Pedretti}, E., {Thureau}, N., {et~al.} 2006, in Society of
  Photo-Optical Instrumentation Engineers (SPIE) Conference Series, Vol. 6268,
  Society of Photo-Optical Instrumentation Engineers (SPIE) Conference Series,
  ed. J.~D. {Monnier}, M.~{Sch{\"o}ller}, \& W.~C. {Danchi}, 62681P,
  \dodoi{10.1117/12.671982}

\bibitem[{{Monnier} {et~al.}(2007){Monnier}, {Zhao}, {Pedretti}, {Thureau},
  {Ireland}, {Muirhead}, {Berger}, {Millan-Gabet}, {Van Belle}, {ten
  Brummelaar}, {McAlister}, {Ridgway}, {Turner}, {Sturmann}, {Sturmann}, \&
  {Berger}}]{Monnier2007Sci...317..342M}
{Monnier}, J.~D., {Zhao}, M., {Pedretti}, E., {et~al.} 2007, Science, 317, 342,
  \dodoi{10.1126/science.1143205}

\bibitem[{{Monnier} {et~al.}(2012){Monnier}, {Che}, {Zhao}, {Ekstr{\"o}m},
  {Maestro}, {Aufdenberg}, {Baron}, {Georgy}, {Kraus}, {McAlister}, {Pedretti},
  {Ridgway}, {Sturmann}, {Sturmann}, {ten Brummelaar}, {Thureau}, {Turner}, \&
  {Tuthill}}]{monnier2012}
{Monnier}, J.~D., {Che}, X., {Zhao}, M., {et~al.} 2012, \apjl, 761, L3,
  \dodoi{10.1088/2041-8205/761/1/L3}

\bibitem[{{Monnier} {et~al.}(2018){Monnier}, {Le Bouquin}, {Anugu}, {Kraus},
  {Setterholm}, {Ennis}, {Lanthermann}, {Jocou}, \& {ten
  Brummelaar}}]{Monnier2018SPIE10701E..22M}
{Monnier}, J.~D., {Le Bouquin}, J.-B., {Anugu}, N., {et~al.} 2018, in Society
  of Photo-Optical Instrumentation Engineers (SPIE) Conference Series, Vol.
  10701, Optical and Infrared Interferometry and Imaging VI, ed. M.~J.
  {Creech-Eakman}, P.~G. {Tuthill}, \& A.~{M{\'e}rand}, 1070122,
  \dodoi{10.1117/12.2312762}

\bibitem[{{Murphy} {et~al.}(2014){Murphy}, {Bedding}, {Shibahashi}, {Kurtz}, \&
  {Kjeldsen}}]{Murphy2014MNRAS.441.2515M}
{Murphy}, S.~J., {Bedding}, T.~R., {Shibahashi}, H., {Kurtz}, D.~W., \&
  {Kjeldsen}, H. 2014, \mnras, 441, 2515, \dodoi{10.1093/mnras/stu765}

\bibitem[{{Murphy} {et~al.}(2018){Murphy}, {Moe}, {Kurtz}, {Bedding},
  {Shibahashi}, \& {Boffin}}]{Murphy2018MNRAS.474.4322M}
{Murphy}, S.~J., {Moe}, M., {Kurtz}, D.~W., {et~al.} 2018, \mnras, 474, 4322,
  \dodoi{10.1093/mnras/stx3049}

\bibitem[{{Offner} {et~al.}(2010){Offner}, {Kratter}, {Matzner}, {Krumholz}, \&
  {Klein}}]{Offner2010}
{Offner}, S.~S.~R., {Kratter}, K.~M., {Matzner}, C.~D., {Krumholz}, M.~R., \&
  {Klein}, R.~I. 2010, \apj, 725, 1485, \dodoi{10.1088/0004-637X/725/2/1485}

\bibitem[{{Paxton} {et~al.}(2011){Paxton}, {Bildsten}, {Dotter}, {Herwig},
  {Lesaffre}, \& {Timmes}}]{Paxton2011ApJS..192....3P}
{Paxton}, B., {Bildsten}, L., {Dotter}, A., {et~al.} 2011, \apjs, 192, 3,
  \dodoi{10.1088/0067-0049/192/1/3}

\bibitem[{{Paxton} {et~al.}(2013){Paxton}, {Cantiello}, {Arras}, {Bildsten},
  {Brown}, {Dotter}, {Mankovich}, {Montgomery}, {Stello}, {Timmes}, \&
  {Townsend}}]{Paxton2013ApJS..208....4P}
{Paxton}, B., {Cantiello}, M., {Arras}, P., {et~al.} 2013, \apjs, 208, 4,
  \dodoi{10.1088/0067-0049/208/1/4}

\bibitem[{{Paxton} {et~al.}(2015){Paxton}, {Marchant}, {Schwab}, {Bauer},
  {Bildsten}, {Cantiello}, {Dessart}, {Farmer}, {Hu}, {Langer}, {Townsend},
  {Townsley}, \& {Timmes}}]{Paxton2015ApJS..220...15P}
{Paxton}, B., {Marchant}, P., {Schwab}, J., {et~al.} 2015, \apjs, 220, 15,
  \dodoi{10.1088/0067-0049/220/1/15}

\bibitem[{{Paxton} {et~al.}(2018){Paxton}, {Schwab}, {Bauer}, {Bildsten},
  {Blinnikov}, {Duffell}, {Farmer}, {Goldberg}, {Marchant}, {Sorokina},
  {Thoul}, {Townsend}, \& {Timmes}}]{Paxton2018ApJS..234...34P}
{Paxton}, B., {Schwab}, J., {Bauer}, E.~B., {et~al.} 2018, \apjs, 234, 34,
  \dodoi{10.3847/1538-4365/aaa5a8}

\bibitem[{{Pourbaix}(2000)}]{Pourbaix2000A&AS..145..215P}
{Pourbaix}, D. 2000, \aaps, 145, 215, \dodoi{10.1051/aas:2000237}

\bibitem[{{Qin} {et~al.}(2019){Qin}, {Luo}, {Hou}, {Li}, {Zhang}, {Wang},
  {Wang}, {Kong}, \& {Han}}]{Qin2019ApJS..242...13Q}
{Qin}, L., {Luo}, A.~L., {Hou}, W., {et~al.} 2019, \apjs, 242, 13,
  \dodoi{10.3847/1538-4365/ab17d8}

\bibitem[{{Raghavan} {et~al.}(2010){Raghavan}, {McAlister}, {Henry}, {Latham},
  {Marcy}, {Mason}, {Gies}, {White}, \& {ten Brummelaar}}]{Raghavan2010}
{Raghavan}, D., {McAlister}, H.~A., {Henry}, T.~J., {et~al.} 2010, \apjs, 190,
  1, \dodoi{10.1088/0067-0049/190/1/1}

\bibitem[{{Reid} {et~al.}(2006){Reid}, {Lewitus}, {Allen}, {Cruz}, \&
  {Burgasser}}]{Reid2006}
{Reid}, I.~N., {Lewitus}, E., {Allen}, P.~R., {Cruz}, K.~L., \& {Burgasser},
  A.~J. 2006, \aj, 132, 891, \dodoi{10.1086/505626}

\bibitem[{{Rizzuto} {et~al.}(2013){Rizzuto}, {Ireland}, {Robertson}, {Kok},
  {Tuthill}, {Warrington}, {Haubois}, {Tango}, {Norris}, {ten Brummelaar},
  {Kraus}, {Jacob}, \& {Laliberte-Houdeville}}]{Rizzuto2013MNRAS.436.1694R}
{Rizzuto}, A.~C., {Ireland}, M.~J., {Robertson}, J.~G., {et~al.} 2013, \mnras,
  436, 1694, \dodoi{10.1093/mnras/stt1690}

\bibitem[{{Rogers} {et~al.}(1974){Rogers}, {Hinteregger}, {Whitney},
  {Counselman}, {Shapiro}, {Wittels}, {Klemperer}, {Warnock}, {Clark},
  {Hutton}, {Marandino}, {Ronnang}, {Rydbeck}, \&
  {Niell}}]{Rogers1974ApJ...193..293R}
{Rogers}, A.~E.~E., {Hinteregger}, H.~F., {Whitney}, A.~R., {et~al.} 1974,
  \apj, 193, 293, \dodoi{10.1086/153162}

\bibitem[{{Sana} {et~al.}(2012){Sana}, {de Mink}, {de Koter}, {Langer},
  {Evans}, {Gieles}, {Gosset}, {Izzard}, {Le Bouquin}, \&
  {Schneider}}]{Sana2012}
{Sana}, H., {de Mink}, S.~E., {de Koter}, A., {et~al.} 2012, Science, 337, 444,
  \dodoi{10.1126/science.1223344}

\bibitem[{{Swihart} {et~al.}(2017){Swihart}, {Garcia}, {Stassun}, {van Belle},
  {Mutterspaugh}, \& {Elias}}]{Swihart2017AJ....153...16S}
{Swihart}, S.~J., {Garcia}, E.~V., {Stassun}, K.~G., {et~al.} 2017, \aj, 153,
  16, \dodoi{10.3847/1538-3881/153/1/16}

\bibitem[{{ten Brummelaar} {et~al.}(2005){ten Brummelaar}, {McAlister},
  {Ridgway}, {Bagnuolo}, {Turner}, {Sturmann}, {Sturmann}, {Berger}, {Ogden},
  {Cadman}, {Hartkopf}, {Hopper}, \&
  {Shure}}]{tenBrummelaar2005ApJ...628..453T}
{ten Brummelaar}, T.~A., {McAlister}, H.~A., {Ridgway}, S.~T., {et~al.} 2005,
  \apj, 628, 453, \dodoi{10.1086/430729}

\bibitem[{{Tokovinin} \& {Moe}(2020)}]{Tokovinin2020MNRAS.491.5158T}
{Tokovinin}, A., \& {Moe}, M. 2020, \mnras, 491, 5158,
  \dodoi{10.1093/mnras/stz3299}

\bibitem[{{Webbink}(1984)}]{Webbink1984ApJ...277..355W}
{Webbink}, R.~F. 1984, \apj, 277, 355, \dodoi{10.1086/161701}

\bibitem[{{Whelan} \& {Iben}(1973)}]{Whelan1973ApJ...186.1007W}
{Whelan}, J., \& {Iben}, Icko, J. 1973, \apj, 186, 1007, \dodoi{10.1086/152565}

\bibitem[{{Winters} {et~al.}(2019){Winters}, {Henry}, {Jao}, {Subasavage},
  {Chatelain}, {Slatten}, {Riedel}, {Silverstein}, \& {Payne}}]{Winters2019}
{Winters}, J.~G., {Henry}, T.~J., {Jao}, W.-C., {et~al.} 2019, arXiv e-prints.
\newblock \doarXiv{1901.06364}

\bibitem[{{Zhao} {et~al.}(2009){Zhao}, {Monnier}, {Pedretti}, {Thureau},
  {M{\'e}rand}, {ten Brummelaar}, {McAlister}, {Ridgway}, {Turner}, {Sturmann},
  {Sturmann}, {Goldfinger}, \& {Farrington}}]{zhao2009ApJ...701..209Z}
{Zhao}, M., {Monnier}, J.~D., {Pedretti}, E., {et~al.} 2009, \apj, 701, 209,
  \dodoi{10.1088/0004-637X/701/1/209}

\end{thebibliography}
\bibliographystyle{aasjournal}

\end{document}